\documentclass[a4paper,fleqn]{cas-dc}

\usepackage{amsmath}  
\usepackage{mathrsfs}  

\usepackage{graphicx}
\usepackage{subcaption}
\usepackage{booktabs}
\usepackage{url}
\usepackage{siunitx}
\usepackage{silence}
\usepackage{mathtools}

\usepackage[boxed,lined,linesnumbered,ruled,vlined]{algorithm2e}  
\usepackage[numbers]{natbib}
\usepackage[english]{babel}



\SetKwInput{KwInput}{Input}                
\SetKwInput{KwOutput}{Output}              

\def\tsc#1{\csdef{#1}{\textsc{\lowercase{#1}}\xspace}}
\tsc{WGM}
\tsc{QE}
\tsc{EP}
\tsc{PMS}
\tsc{BEC}
\tsc{DE}

\begin{document}
\let\WriteBookmarks\relax
\def\floatpagepagefraction{1}
\def\textpagefraction{.001}

\shorttitle{Emerging Threats in Deep Learning-Based Autonomous Driving}

\shortauthors{Hui Cao et~al.}

\title [mode = title]{Emerging Threats in Deep Learning-Based Autonomous Driving: A Comprehensive Survey}                      

\tnotetext[1]{This document is the results of the research project funded by the National Natural Science Foundation of China: Research on secure data management mechanism of new college entrance examination comprehensive quality evaluation: A security enhancement of block chain empowerment(No. 72204077)}

\tnotetext[2]{This document is the results of the research project funded bt the Hubei Natural Science Foundation project:Research on the data security management mechanism of new College Entrance Examination comprehensive quality evaluation based on blockchain.(No. 2021CFB470)}

%

                        
\author[1]{Cao Hui}[style=chinese]                        

\ead{cao-hui@whu.edu.cn}
\ead[url]{https://scholar.google.com/citations?user=1XoXUTYAAAAJ&hl=en}

\author[1]{Zou Wenlong}[style=chinese]


\author%
[2]
{Wang Yinkun}[style=chinese]

\author%
[3]
{Song Ting}[style=chinese]

\author
[1]{Liu Mengjun}[style=chinese]
    \cormark[1]
    \ead{lmj_whu@163.com}


\address[1]{School of Education,Hubei University,Youyi Road No.368,Wuhan,430062,Hubei,P.R.China}


\address[2]{Institute of Information Engineering,Chinese Academy of Sciences, Beijing,100093,P.R.China}
    
\address[3]{School of Foreign Languages,Hubei University,Youyi Road,No.368, Wuhan,430062,Hubei,P.R.China}
    

\cortext[cor1]{Corresponding author}
\cortext[cor2]{First two author equal contribution}



\begin{abstract}
Since the 2004 DARPA Grand Challenge, the autonomous driving technology has witnessed nearly two decades of rapid development. Particularly, in recent years, with the application of new sensors and deep learning technologies extending to the autonomous field, the development of autonomous driving technology has continued to make breakthroughs. Thus, many carmakers and high-tech giants dedicated to  research and system development of autonomous driving. However, as the foundation of autonomous driving, the deep learning technology faces many new security risks. The academic community has proposed deep learning countermeasures against the adversarial examples and AI backdoor, and has introduced them into the autonomous driving field for verification. Deep learning security matters to autonomous driving system security, and then matters to personal safety, which is an issue that deserves attention and research.This paper provides an summary of the concepts, developments and recent research in deep learning security technologies in autonomous driving. Firstly, we briefly introduce the deep learning framework and pipeline in the autonomous driving system, which mainly include the deep learning technologies and algorithms commonly used in this field. Moreover, we focus on the potential security threats of the deep learning based autonomous driving system in each functional layer in turn. We reviews the development of deep learning attack technologies to autonomous driving, investigates the State-of-the-Art algorithms, and reveals the potential risks. At last, we provides an outlook on deep learning security in the autonomous driving field and proposes recommendations for building a safe and trustworthy autonomous driving system.
\end{abstract}



\begin{keywords}
 Trustworthy AI\\
 Deep Learning\\
 Artificial Intelligence\\
 Autonomous Driving\\
 Cyber Security\\
 Adversarial Examples 
\end{keywords}

\maketitle

\section{Introduction}

Research about Autonomous Land Vehicles (ALVs) began as early as 1980s with funding from the US Department of Defense(DoD). In the 21st century, DARPA conducted the Grand Challenge that launched a new generation of autonomous driving. The development of artificial  intelligence(AI) technology is driving the rapid progress of autonomous vehicles with an increasing expectation from the public. Currently, many traditional carmakers, such as universal Motors, Toyota, Volvo, BMW and Audi have carried out researches into the autonomous driving system. On another hand, not to be outdone, most of high-tech giants, Google Waymo, Tesla, Baidu and Huawei, devoted themselves to autonomous driving technology. Along with artificial intelligence technology, autonomous driving has seen rapid development and is expected to enter the practical stage.

However, security is a major concern in the application of the autonomous driving system, because there are new types of security risks associated with autonomous driving system that depends heavily on deep learning. On the one hand, from the perspective of technical threat on AI security and privacy protection, new countermeasures have been proposed successively, including adversarial examples \cite{Szegedy2013intriguing,Goodfellow2014explaining}, data poisoning and AI Backdoor\cite{Gu2017badnets}, model extraction\cite{Tramer2016stealing}, model inversion\cite{Fredrikson2015model}, and membership privacy inference\cite{Shokri2017membership}. On the another hand, from the social trust perspective of AI, issues about fairness, AI abuse, environment, compliance, and ethic, have also received attention and research. Currently, there is some literature\cite{He2020towards,Ji2021,Xu2020adversarial,Li2020adversarial,Zhang2019adversarial,Yuan2019adversarial,Goldblum2020data} summarized AI security threats in the general environment. Different from that, this paper focuses on the environment of autonomous driving system, it reveals the new security risks posed by AI technologies bringing new security challenges to autonomous driving. Unlike other applications of deep learning, the autonomous driving system is a more complex AI architecture consisting of dozens of functional modules, and different environment modules with different characteristics, raising different requirements for AI security attack and mitigation techniques, including:
\begin{itemize}

\item \textbf{Physical world requirements.} AI threats of autonomous driving system should be able to take effect in the real physical world, and not only in the digital world and computer simulation systems. Techniques specific to adversarial examples attacks in the physical world are the focus of this paper.

\item \textbf{Robustness requirements.} The environment is uncertain and often varies to a large extent in autonomous driving. On the one hand, the images collected under different weather, light and other natural conditions can vary; on the other hand, changes in a long distance and large angle range also make image acquisition highly variable due to the high-speed movement of vehicles. Therefore, AI threats need to be able to take effect continuously and stably under a variety of conditions, which raises very high demands on the robustness of attacks, such as adversarial examples and AI backdoor. This paper focuses on robustness enhancement methods.

\item \textbf{Fusion environment requirements.} Autonomous driving system often employs multi-modal fusion sensing techniques that combine different types of information from multiple RGB cameras, LiDAR, RaDAR, etc., to sense the fused images. The autonomous driving environment requires that adversarial examples countermeasures and other related threat technologies can be stabilized to remain in effect in the fused environment. Artificial intelligence threats in multi-modal fusion environments are also the focus of this paper.

\end{itemize}
Due to the above concerns and requirements, AI safety technologies in the field of autonomous driving have continued to develop, and some research results and breakthroughs have been achieved. This paper introduces the latest research progress relevant to unique technologies and reveals the AI security risks in autonomous driving systems. This paper faces the above challenges of autonomous driving systems, rather than in the general environment. Section 1 briefly introduces the infrastructure and key technologies of AI in autonomous driving; section 2 offers a glimpse of the AI risks in the sensor layer; section 3 comprehensively reviewed the AI risk in the perception layer, introduced the idea and detail of important algorithms;  section 4  provides the potential deep leaning risk and attack technology in decision layer in autonomous driving; section 5 focus on new threat of V2X that based on federation learning in the future; section 6 gives a summary and outlook. 

\subsection{Basic Concepts of Autonomous Driving}

Essentially, autonomous driving is making driving decisions through artificial intelligence techniques or other automated decision-making methods. According to the Society of Automotive Engineers (SAE) standard J3016\cite{Sae2014}, autonomous driving can be categorized into the following classes.

\begin{itemize}

\item \textbf{L0 -- No Driving Automation}: driving is carried out entirely by a person, but warnings and system assistance are available during the journey.

\item \textbf{L1 -- Driver Assistance}: based on the perception of the driving environment, only a single aspect of automation, which system operates the steering wheel or acceleration and deceleration assists the driver with ADAS, while other driving operations are performed by the human driver.

\item \textbf{L2 -- Partial Driving Automation}: based on the perception of the driving environment, the system operates both the steering wheel and acceleration or deceleration. However, it requires a human driver to remain constantly alert and ready to take full control with little or no warning.

\item \textbf{L3 -- Conditional Driving Automation}: based on the perception of the driving environment, autonomous driving system can perform all driving operations under the supervision of a human driver.

\item \textbf{L4 -- High Driving Automation}: under certain environmental conditions, autonomous driving system can perform all driving operations unsupervised.

\item \textbf{L5 -- Full Driving Automation}: the autonomous driving system can perform all driving operations unsupervised in all environmental conditions.

\end{itemize}
For autonomous driving system, there are different views and concepts, as well as different development and evolutionary routes. One is focus on intelligentization and cyberization of vehicle components, mainly researching on sensors, in-vehicle communication, vehicle-to-everything (V2X), and et al, which main participant by traditional car-makers. The other is focus on autonomous diving decisions, mainly researching artificial intelligence and autonomous driving, and the main participants include: UC Berkeley, Google WayMo, Baidu, Apollo, Intel Carla, NVIDIA and other artificial intelligence companies. However, whether it starts from the vehicle moving towards AI or the other way round, automated driving decision is the core mission in autonomous driving, and safety based on AI driving decisions making is a necessary prerequisite for the safety of autonomous driving system. The higher the level of autonomous driving, the higher the reliance on AI technology represented by deep learning, which lead to higher the requirements for the safety and robustness of deep learning itself.

\subsection{Architecture of Autonomous Driving System}

In terms of the autonomous driving architecture and machine learning technologies based, autonomous driving systems can be divided into end-to-end (E2E) and modular architectures.

\begin{figure}[htbp]
\centering
\begin{subfigure}{0.5\textwidth}
\centering
\includegraphics[width=7.5cm]{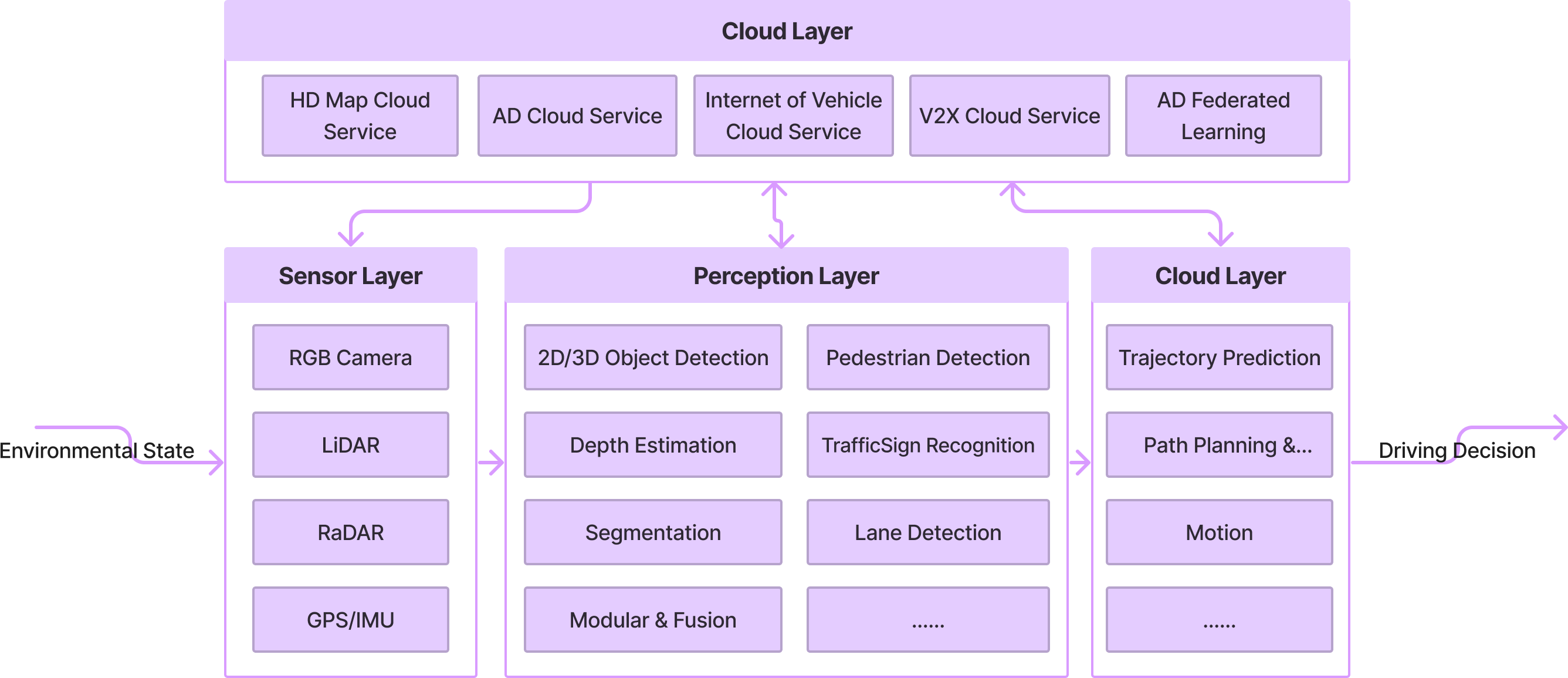}
\caption{Modular Autonomous Driving Framework}
\end{subfigure}\vspace{0.5cm}
\begin{subfigure}{0.5\textwidth}
\centering
\includegraphics[width=7.5cm]{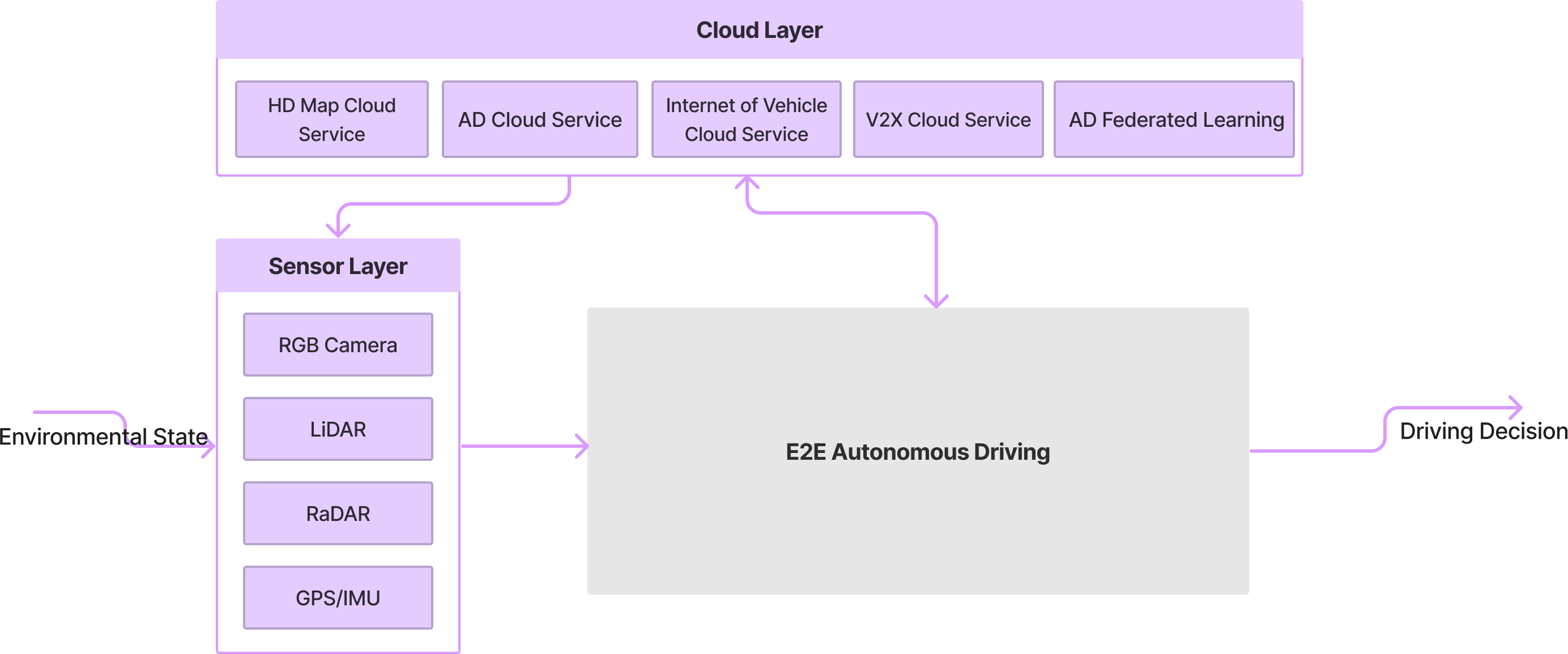}
\caption{E2E Autonomous Driving  Framework}
\end{subfigure}
\caption{Autonomous Driving Framework}
\end{figure}

\begin{itemize}

\item The modular autonomous driving system divides an individual set of autonomous driving functions into several parts, each of which is completed by one or a group of artificial intelligence models, usually including: positioning and projecting, target recognition, trajectory prediction, road planning \& driving decision making, vehicle control, and other functions. These functional modules contain the sensing layer, the perception layer, the decision layer and the vehicle networking layer.

\item The End-to-End autonomous driving system often consists of a large number of complex judgment functions in driving decisions performed by one or a group of artificial intelligence models that make the final driving decision based on the environment and cloud inputs.

\end{itemize}

\subsection{Sensing Layer}

The sensing layer includes a variety of sensors that collect information about the environment for the autonomous driving system. Common sensors used in autonomous driving vehicles compromise RGB cameras, LiDAR (Light Detection and Ranging), RaDAR (Radio Waves to Determine the Distance), GPS, and ultrasonic sensors. Here are the characteristics of different sensors:

\begin{itemize}

\item The advantages of RGB cameras are: 1) lower cost, and 2) relatively mature recognition technology; their limitation is that the distance is dependent on estimation.

\item The advantage of LiDAR is that it is accurate; its limitation is that it is susceptible to interference from the weather.

\item The advantage of radar is that it is relatively immune to weather interference; its limitation is that it has insufficient imaging capability.

\end{itemize}

There are a number of existing works that provide a detailed comparison of sensors for autonomous driving vehicles, which will not be the emphasis of this paper. There are some survey papers related to the sensing layer\cite{Ren2019security,Huang2020survey,Deng2021deep}.

Most companies have chosen autonomous driving technology solutions that multi-modal fusion, while some have chosen solutions that rely primarily on RGB cameras. However, it needs to be emphasized that, regardless of the choice of sensor configuration solution, the various advanced sensors only fulfill the function of raw information collection and do not replace the key role played by artificial intelligence in the perception and decision-making of autonomous driving system, and are equally unable to avoid the new safety risks posed by AI.

\subsection{Perception Layer}

The perception layer perceives and identifies things like object perceiving and identification, segmentation, depth estimation and localization, which are based on the vehicle's state and road information collected by the sensors in the sensor layer. The commonly used techniques are given as follows, which include 2D object recognition, 3D object recognition, multi-modal fusion, trajectory prediction, and so on.

\begin{figure}[htbp]
\centering
\includegraphics[width=\linewidth]{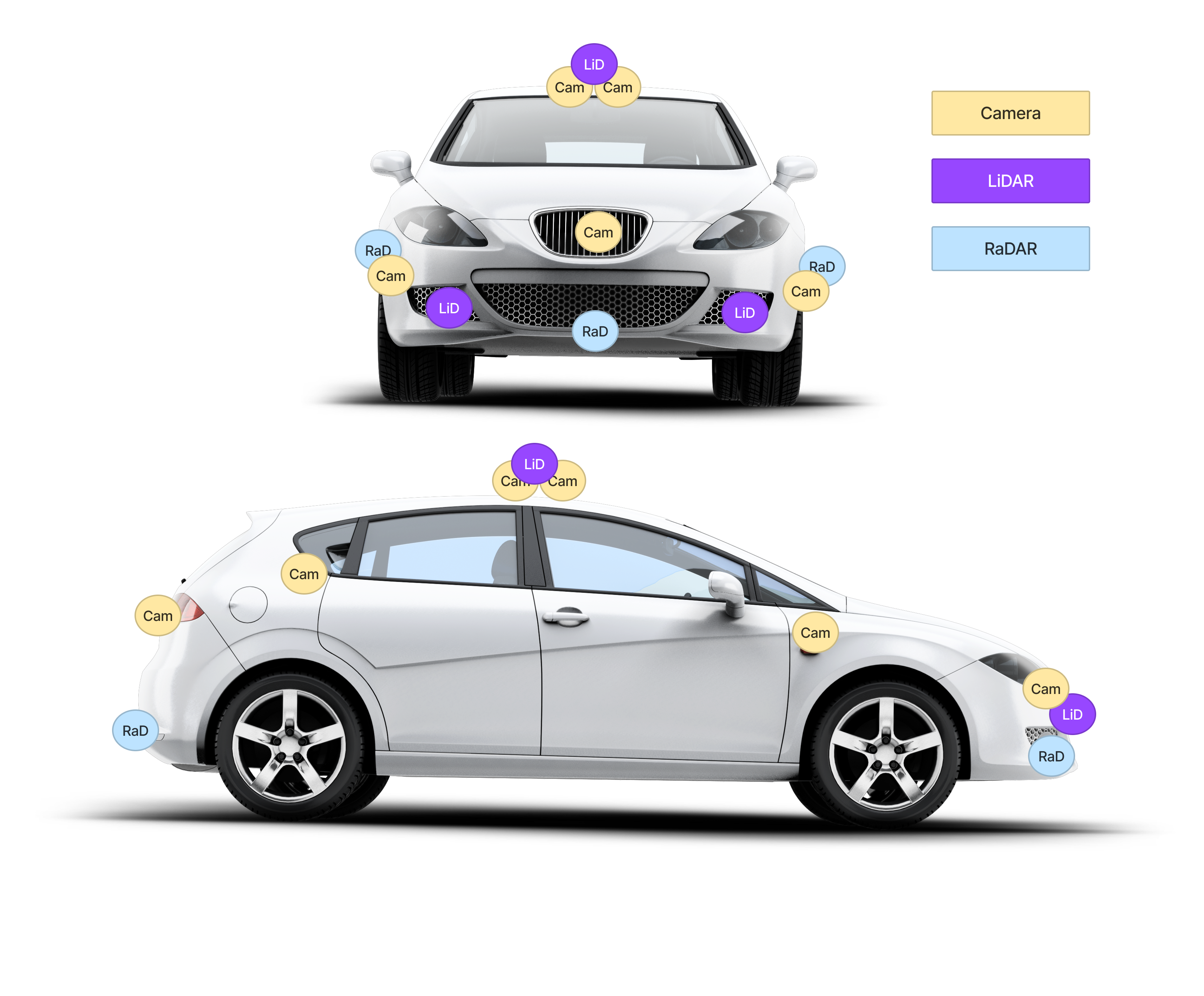} 
\caption{Sensor in Autonomous Driving }\label{Fig1.2}
\end{figure}

\begin{itemize}

\item  \textbf{2D Objection Recognition} is based on a flat image to identify the presence or absence of a specific target in the image and locate it. technologically, 2D object recognition can be divided into two classifications: two-stage objection recognition algorithms and one-stage objection recognition algorithms. The two-stage algorithms first find a series of region proposals, and then classify the objects in the proposals by Convolutional Neural Networks(CNN). Commonly used two-stage algorithms include FasterRCNN\cite{Ren2015faster} and MaskRCNN\cite{He2017mask} characterized by relatively high accuracy and high consumption. One-stage algorithms do not generate a separate region proposal but return to the predicted class and location of the target directly. Commonly used one-stage algorithms include: SSD\cite{Liu2016ssd}and Yolo v3\cite{Redmon2018yolov3}. In 2017, Lin et al.\cite{Lin2017focal} proposed a new loss function - "Focal Loss", which can significantly improve the accuracy of dense target recognition, and this technique was first applied to the field of face recognition. It is now applied to many target recognition fields, among which, in 2021, Yosuke Shinya et al.\cite{Shinya2021usb} proposed UniverseNet, a target detection algorithm that applies  Focal Loss, which can achieve better results in dense target and small target scenarios. A detailed comparison of current mainstream 2D target recognition techniques can be found in references\cite{Ren2019security, Huang2020survey, Deng2021deep}

\item  \textbf{Multi-Modal Fusion}. A single type of sensor cannot capture all of the environmental information needed to support autonomous driving, while autonomous driving systems require information from several types and a large number of sensors to make integrated decisions, which leads us to make multi-modal fusion. Depending on occurred times\cite{Feng2020deep}, the fusion can be divided into three modes: \textbf{pre-fusion}, \textbf{post-fusion},  and \textbf{deep fusion}. Pre-fusion combines the data collected by all types of sensors and then makes a comprehensive decision. Post-fusion to make decisions on the data collected by different sensors and then aggregate the sub-decisions. Deep fusion constitute by the fusion of data, features and decision integration, and can be subdivided into five types: \textbf{data in data out}, \textbf{data in feature out}, \textbf{feature in feature out}, \textbf{feature in decision out}, and \textbf{decision in decision out}\cite{Dasarathy1997sensor,Fayyad2020deep}. An in-depth analysis and comparison of the various integration methods can be found in the literature.\cite{Fayyad2020deep,Wang2021multi,Yeong2021sensor,Fayyad2020deep}

\begin{figure}[htbp]
\centering
\begin{subfigure}{7cm}
\centering
\includegraphics[width=7cm]{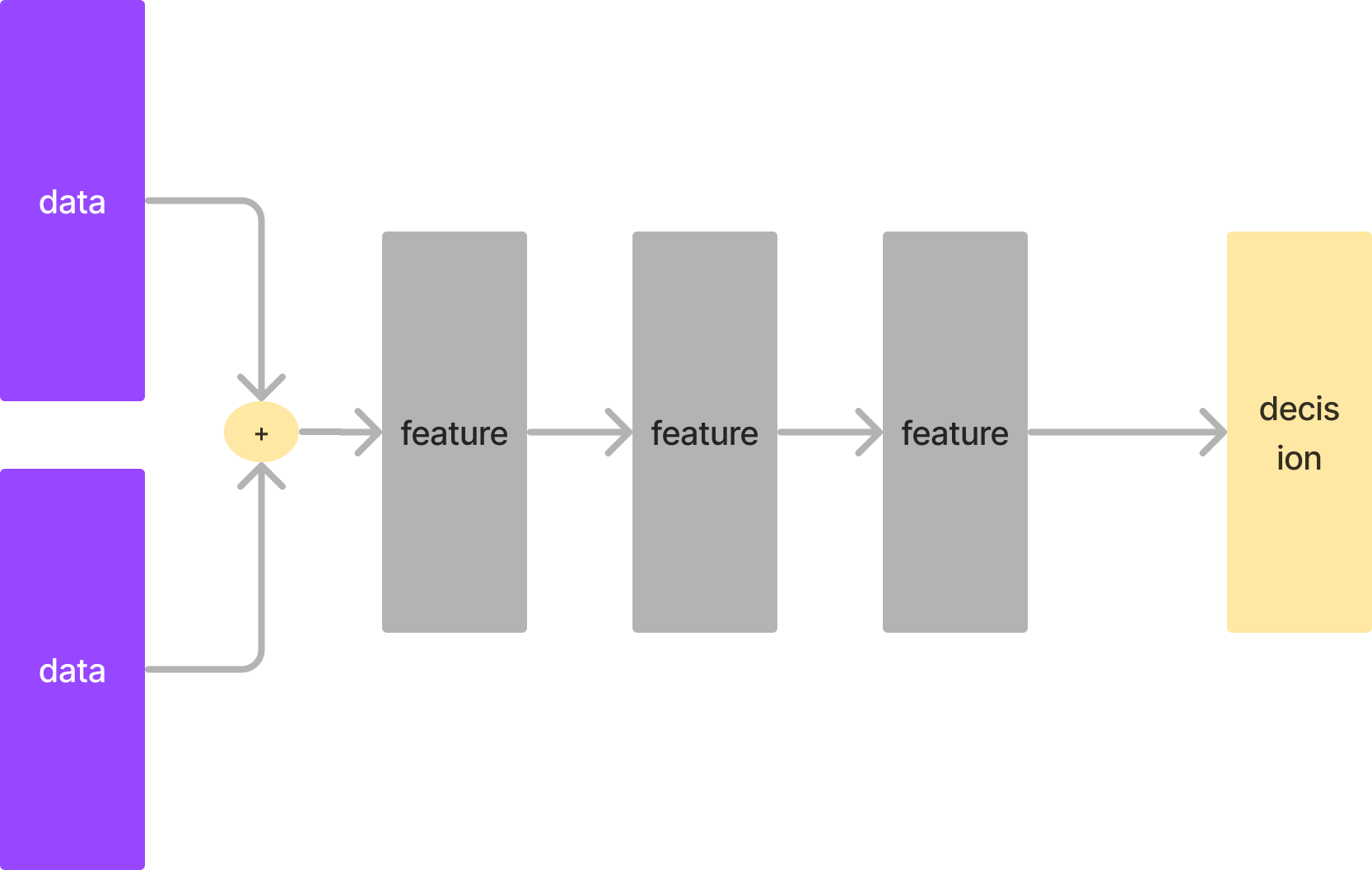}
\subcaption{Pre-fusion}
\end{subfigure}\vspace{0.75cm}

\begin{subfigure}{7cm}
\centering
\includegraphics[width=7cm]{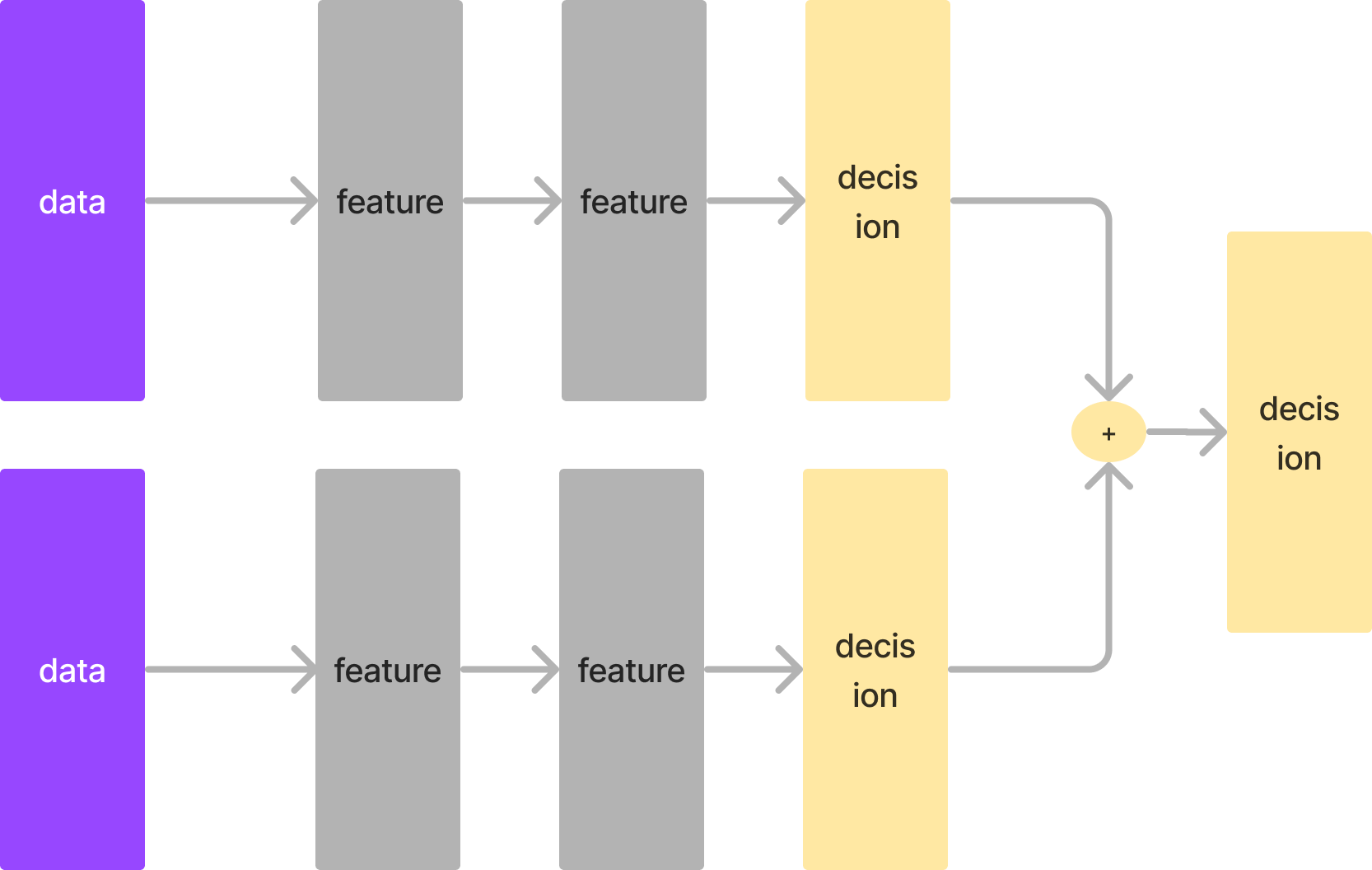}
\subcaption{Post-fusion}
\end{subfigure}\vspace{0.75cm}

\begin{subfigure}{7cm}
\centering
\includegraphics[width=7cm]{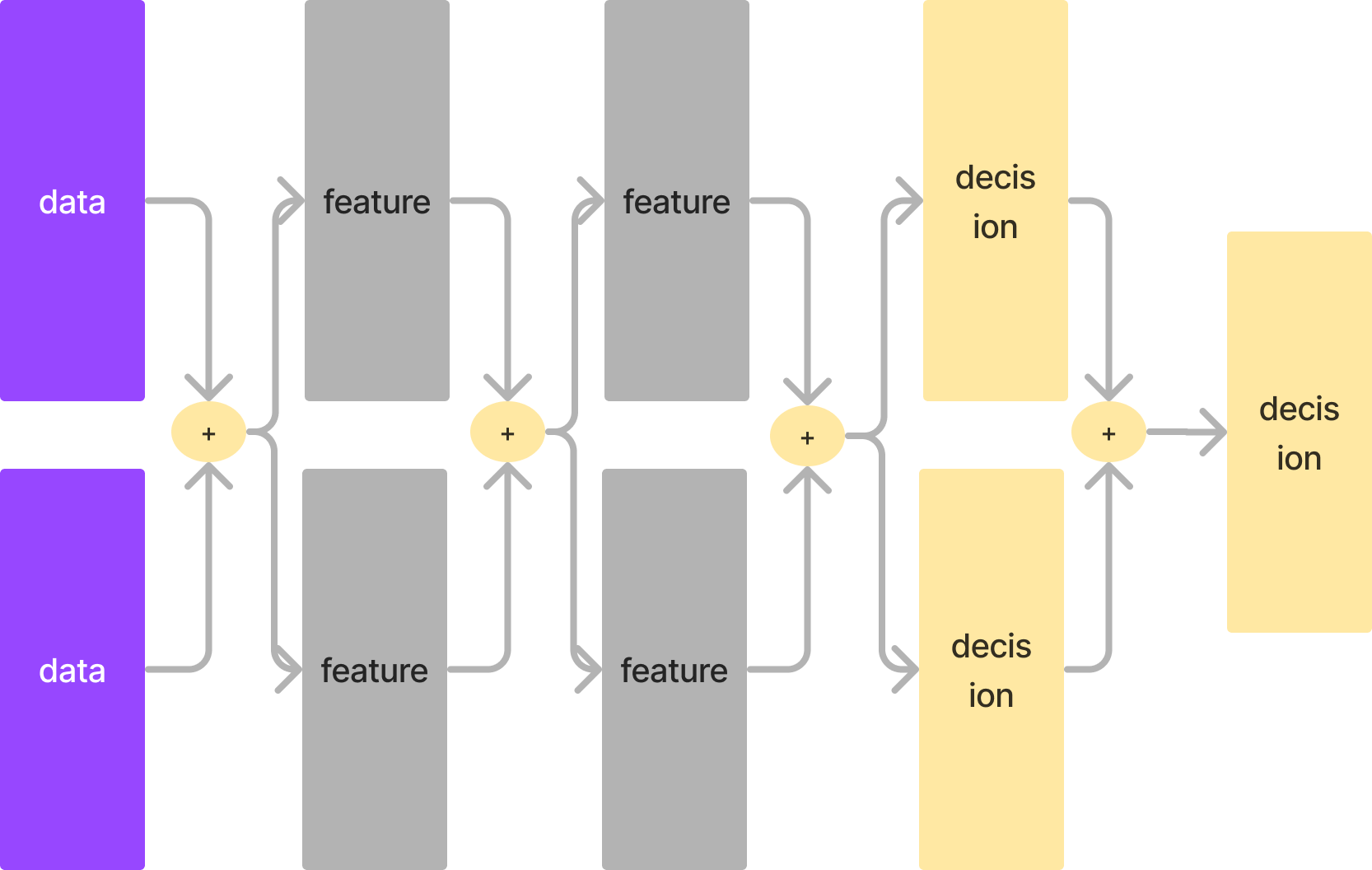}
\subcaption{Deep fusion}
\end{subfigure}
\caption{Fusion of Autonomous Driving}
\end{figure}

\item  \textbf{3D Objection Detection and Segmentation}. Because 2D images have no depth information that is needed in autonomous driving, such as path planning and collision avoidance in autonomous driving, therefore 3D objection detection plays a key role. Classified by the detected information, 3D target detection has 3 bases: 2D image, 3D point cloud map and multi-modal fusion image. Among them, 3D target detection based on 2D images often uses 3D target matching and depth estimation to estimate the 3D target bounding box for targets in 2D images using algorithms like Mono3D\cite{Chen2016monocular}, 3DVP\cite{Xiang2015data}, Deepmanta\cite{Chabot2017deep}, and SVGA-Net\cite{He2020svga}. 3D target recognition based on 3D point cloud maps is the recognition of targets in the images with 3D information and marks the target outline. Commonly used algorithms include: VeloFCN\cite{Li2016vehicle} , BirdNet\cite{Beltran2018birdnet}, 3DFCN\cite{Li20173d} , PointNet++\cite{Qi2017pointnet++} and VoxelNet\cite{Zhou2018voxelnet}. 3D target detection based on multi-modal integration images is to use different integration modes to identify 3D targets. Commonly used algorithms include: MV3D\cite{Chen2017multi}, AVOD\cite{Ku2018joint}, and F-PointNet\cite{Qi2018frustum}. A comparison and in-depth study of various 3D target detection algorithms can be found in the literature\cite{Fernandes2021point, Qian20223d}.

\end{itemize}
Other deep learning research directions in the perception layer include Pedestrian Detection, Lane Detection, Traffic Sign Recognition, Pedestrian Attribute Recognition, Fast Vehicle Detection, Pedestrian Density Estimation, Plate Recognition, etc. There is detail on the leaderboard\cite{SOTA}. 

\subsection{Decision-Making Layer}

Driving decision-making is the core of autonomous driving, and machine learning methods are often used, with two technical routes available: Imitation Learning and Reinforcement Learning.
\begin{itemize}

\item   \textbf{Imitation Learning}. Imitation learning refers to the learning behavior of agents who acquire the ability to perform a specific task by observing and imitating the behavior of human experts\cite{Hussein2017imitation}. Imitation learning has been successful in the field of autonomous driving\cite{Pan2020imitation} Imitation learning tends to collect a large amount of environmental state $S_{i}$ (environmental data collected by various sensors, including 3D point cloud maps, RGB images, etc.) as features and record the actions performed by the human experts at the same time. $A_{i}$ is used as a label to form a training data set $D:{(s1,a1),(s2,a2),(s3,a3),...}$. Using specific imitation learning algorithms, artificial intelligence models are trained and used to make future driving decisions. The famous imitation learning methods include the E2E autonomous driving algorithm based on conditional imitation learning \cite{Codevilla2018end}, and the ChauffeurNet\cite{Bansal2018chauffeurnet}. 

\item   \textbf{Deep Reinforcement Learning}. Deep reinforcement learning simulates the self-learning model of organisms in nature. To be concrete, an agent monitors its own behavior and the resulting environmental changes, sets the reward value for different changes, and then continuously optimizes the model and its own behavior based on this. In 2013, Mnih et al.\cite{Mnih2015human} combined deep learning with reinforcement learning and proposed the Deep Q Learning(DQN) method. DQN is based on a set of Q values in a reward table. The system's driving status $S_{i}$ and the driving operation $a_{i}$ to obtain the corresponding reward value $r_{i}$, which automatically generates training data  $D:{((s1,a1),r1),((s2,a2),r2),((s3,a3),r3),...}$. The reinforcement learning model is then trained by specific algorithms, while reinforcement learning is supplemented with current operational data to continuously optimize the model. Nowadays, deep reinforcement learning has been rapidly developed and widely used, with subsequently emerged Deep Recurrent Q Networks (DRQNs)\cite{Hausknecht2015deep}, attention mechanism deep recurrent Q networks\cite{Sorokin2015deep}, asynchronous/synchronous dominant actor-critic (A3C/A2C)\cite{Mnih2016asynchronous}, and reinforcement learning for unsupervised and unassisted tasks\cite{Jaderberg2016reinforcement}, which are widely used in e-Sports, health \& medicine, recommendation system and other fields. There are some surveys of deep reinforcement learning\cite{Sutton2018reinforcement,Arulkumaran2017deep}. 

A variety of deep reinforcement learning frameworks and algorithms are widely used in the field of autonomous driving vehicles. For example, Feng et al.\cite{Feng2019autonomous}, Alizadeh et al.\cite{Alizadeh2019automated}, Mirchevska et al.\cite{Mirchevska2018high}, and Quek et al.\cite{Quek2021deep} apply deep reinforcement learning techniques to driving decisions; Holen et al.\cite{Holen2020road}  use deep reinforcement learning for autonomous driving roadway recognition; Feng et al.\cite{Feng2020environmental} utilize deep reinforcement learning techniques for traffic light optimization control. Some researchers have also proposed an  autonomous driving solution with the fusion of imitation learning and reinforcement learning\cite{Kiran2021deep,Aradi2020survey}.

\end{itemize}

\subsection{Vehicle Networks}

With the development of communications and AI technology, vehicle networks are increasingly playing an important role in autonomous driving, especially the vehicle networks construction, which supports a distributed AI model and provides a novel type of AI technology in autonomous driving, while also bringing new security risks.
\begin{itemize}

\item  Vehicle-to-Everything (V2X). V2X is a multi-layered network system designed to enhance collaboration between pedestrians, vehicles and transport infrastructure. It is universally composed of Vehicle-to-Vehicle (V2V) networks, Vehicle-to-Infrastructure (V2I) networks, Vehicle-to-Pedestrian (V2P) networks and Vehicle-to-Road side units (V2R) networks\cite{Pope2016etsi}. The communication technologies used in the vehicular internet of things can be broadly classified into two categories, Dedicated Short Range Communication (DSRC) and Long-Term Evolution (LTE) cellular communication, called cellular-V2X or C-V2X for short\cite{Papathanassiou2017cellular}. 

\item   Federated Learning. The vehicular internet of things provides the network foundation for distributed artificial intelligence. Federated Learning is a distributed AI framework that replaces sensitive data interactions with model interactions, enabling more efficient and better privacy for knowledge sharing and transition. Based on the V2X, the federated learning can provide distributed and interactive AI services\cite{Ma20195g,Du2020federated,Posner2021federated} for autonomous driving system. This paper focuses on the novel security risks posed by Federated Learning in the vehicular internet of things, and reviews related security technology developments.

\end{itemize}

\subsection{Summary}

We concluded the major AI application used in autonomous driving in Table1

\begin{table*}[h]
\caption{Major Deep Learning-based Tasks in Autonomous Driving}
\label{table1.1}
\begin{center}
\begin{tabular}{ p{2.5cm}<{\centering} | p{5cm}<{\centering} | p{7.5cm}<{\centering}  }
 \hline
 \hline
 \rowcolor{lightgray}
 \textbf{Layer} & \textbf{Task} & \textbf{Major typical deep learning algorithm} \\ 
 \hline
 \hline
 
 \rowcolor{green!50!yellow!10}
 \multirow{2}{*}{Sensor} & 
 3D PointCloud Registration & 
 3DFeat-Net\cite{yew20183dfeat}, FCGF\cite{choy2019fully}, D3Feat-pred\cite{li2020end} \\
 
 & 
 Pre-Fusion & 
 Multi-Frame Fusion\cite{ren2019fusion}, MTF4VT\cite{wang2022multimodal},TransFuser\cite{chitta2022transfuser},DeepFusion\cite{li2022deepfusion} \\
 
\hline
 
\multirow{5}{*}{Perception} &
2D Object Detection &

Fast-RCNN\cite{girshick2015fast}, Faster R-CNN\cite{girshick2015fast}, Mask R-CNN\cite{He2017mask}, D-RFCN\cite{misra2019mish}, Yolov4\cite{qiao2021detectors}, YOLOv7\cite{wang2022yolov7},FD-SwinV2\cite{wei2022contrastive} \\
 
 \rowcolor{green!50!yellow!10}
 & 
 3D Object Detection & 
PointRCNN\cite{shi2019pointrcnn}, PV-RCNN\cite{shi2020pv},Se-SSD\cite{zheng2021se}, GLENet-VR\cite{zhang2022glenet} \\ 

 & 
 Lane Detection & 
SCNN\cite{parashar2017scnn}
LaneATT\cite{tabelini2021keep}
CLRNet\cite{zheng2022clrnet} \\ 
 
  \rowcolor{green!50!yellow!10}
 & 
 Traffic Sign Recognition & 
 CNN with 3 Spatial Transformer\cite{arcos2018deep}, Mask R-CNN with adaptations and augmentations\cite{He2017mask}, TSR-SA\cite{chen2022real} \\ 

 & 
 Fast Vehicle Detection & 
 YOLOv3-tiny\cite{rani2021littleyolo}, LittleYolo-SPP\cite{adarsh2020yolo} \\ 
 
 \rowcolor{green!50!yellow!10}
 & 
 Pedestrian detection & 
SA-FastRCNN\cite{li2017scale},RPN+BF\cite{zhang2016faster},Pedestron\cite{hasan2021generalizable}, \\
 
 & 
 Semantic Segmentation & 
FCN\cite{long2015fully}, PSPNet\cite{zhao2017pyramid}, DRAN\cite{fu2020scene},Swin trasformer\cite{liu2021swin}, ViT-Adapter\cite{chen2022vision} \\
 
  \rowcolor{green!50!yellow!10}
 & 
 Object Tracking & 
 M2-Track\cite{zheng2022beyond},BAT\cite{zheng2021box} \\
 
 & 
 Multiple Object Tracking & 
 QDTrack\cite{chen2022vision}, RetinaTrack\cite{Lu2020retinatrack} \\
 \hline 
 
\rowcolor{green!50!yellow!10}
\multirow{3}{*}{Decision} & 
 Trajectory Prediction  & 
NSP-SFM\cite{yue2022human}, Y-Net\cite{mangalam2021goals},Trajectron++\cite{salzmann2020trajectron++},Social GAN\cite{gupta2018social},SoPhie\cite{sadeghian2019sophie} \\

  & 
Motion Forecasting & 
VI LaneIter\cite{shi2022motion}, Wayformer\cite{nayakanti2022wayformer}\\ 

\rowcolor{green!50!yellow!10}
& 
Deep Reinforcement Learning & 
Deep Q-Learning\cite{mnih2013playing}),Deep recurrent q-learning\cite{Hausknecht2015deep},Deep attention recurrent Q-network\cite{Sorokin2015deep},Double Q-learning\cite{hasselt2010double},A3C/A2C\cite{Mnih2016asynchronous}  \\ 

& 
Imitation Learning & 
Generative adversarial imitation learning\cite{ho2016generative}, Conditional Imitation Learning\cite{Codevilla2018end,Hawke2020urban},  Self-Imitation Learning\cite{oh2018self}, Chauffeurnet\cite{Bansal2018chauffeurnet}   
\\
\hline

\rowcolor{green!50!yellow!10} 
V2X & 
Federated Learning & 
FedAvg\cite{li2020federated} \\

\hline
\hline
 
\end{tabular}
\end{center}
\end{table*}

\section{Emerging Threats in Sensors}

Sensors are foundational part for the autonomous driving system, which provide raw environmental information for autonomous driving decision-making. The security of sensors directly affects the safety of autonomous driving system. We classify attacks against sensors into two categories, where attacks that aim to compromise the usability of the sensing are classified as Jamming Attacks and attacks that aim to compromise the integrity of the information collected by the sensors are classified as Spoofing Attacks.

\subsection{Jamming Attacks}

The Jamming Attack means that attackers take some actions to reduce the quality of data collected by the sensor, even making sensor unavailable. In 2015, Petit et al.\cite{Petit2015remote} attempted a jamming attack on autonomous driving sensors by artificially setting up bright light interference that could "blind" the camera. In 2016, Yan et al.\cite{Yan2016can} experimented with blind attacks on ultrasonic sensors. Similarly, a variety of in-vehicle sensors such as RGB cameras, LiDAR, RaDAR, gyroscopic sensors and GPS sensors could be subject to jamming attacks\cite{Zhang2017dolphinattack,Lim2018autonomous,Son2015rocking,Kar2014detection}.

\subsection{Spoofing Attacks}

The Spoofing Attacks means that attackers injecting fake signals to affect the normal behaviour of the autonomous driving system. In 2015, Petit et al.\cite{Petit2015remote} attempted to send specific spoofed laser signals, causing the LiDAR systems to be misled. Later, Park et al.\cite{Park2016ain} conducted similar experiments on in-vehicle IR sensors. Yan et al.\cite{Yan2016can} worked on gyroscopic sensors and RaDAR. Nassi et al.\cite{Nassi2019mobilbye} conducted combined experiments on RGB cameras, LiDAR and RaDAR. Psiaki et al.\cite{Psiaki2016protecting}, Meng et al.\cite{Meng2019gps} conducted spoofing experiments on GPS for multiple environments.

Currently, most attacks against sensors of vehicle are trend towards physical attack rather than attack on deep learning. In this paper, we only give a general overview, there are more details in the surveys\cite{Deng2021deep, Ren2019security}.

\section{Emerging Threats in Perceptual Layer}

Based on the information captured by various types of sensors in the sensor layer, the perception layer performs recognition and perception. These tasks, such as objective recognition, segmentation, and depth estimation, are often difficult to accomplish through simple computing based on some certain rules. Artificial intelligence is also subject to new types of security threats. For example, attackers could use Adversarial Examples or AI Backdoor attacks, which can mislead to wrong predictions that be controlled by attackers, which leads to dangerous driving decisions. Attackers may also use Model Extraction to obtain the parameters or Hyper-parameters of the AI model, resulting in model leakage and loss of intellectual property. And then attackers would use Model Inversion or Membership Privacy attacks leading to sensitive training data leakage and privacy risks.

Different from some existing surveys that introduce general adversarial examples or AI backdoors in cyberspace, this paper focuses on advanced research in the physical world.  Attacking in the physical world has faced higher demands, especially, on attack constancy, high success rate, and robustness on environmental uncertainties.

\subsection{Physical World Adversarial Examples for Static Objective Detection}

In 2014, researchers discovered that adding a small amount of specific interference, which is imperceptible to human beings, may still cause machine learning to be misled by attackers. This could cause serious security even safety problem, if machine learning be applied to a critical domain.  Such an attack is known as adversarial examples attack and can be formalized as

\begin{equation}
argmin_{x'} \left \| x' -x \right \|_{p}     s.t. f(x') = \hat{y} 
\end{equation}

where $f$ denotes  a machine learning model, $x$ denotes a test example, and $x'$ denotes an adversarial example generated based on the addition of a small amount of interference, with $c$ for the prediction result of the model for a normal example, and $c'$ for the prediction result of the model on the adversarial examples.

After the first work, adversarial examples technique has been widely studied and has seen rapid development. FGSM algorithm proposed by Goodfellow et al.\cite{Goodfellow2014explaining} FGSM works by calculating the gradient of the loss function between the input and target classification and creating a small perturbation in terms of the sign vector coefficient of this gradient as:

\begin{equation}
x_{Adv} = x + \alpha sign(\triangledown_{x} J(x,y) ) 
\end{equation}
 
where $x_{adv}$  denotes the corresponding adversarial example of $x$, $ \alpha $ is a specific constant, $sign()$ is a sign function, $y_{true}$ is the the corresponding true label of $x$, $J()$ denotes the loss function used to train the model, and $\triangledown_{x}$ denotes the the gradient of $x$. The algorithm can be used to add a certain amount of adversarial noise to an image that is normally predicted as a panda, so that the machine learning model can predicted it as a gibbon with a high confidence. This algorithm could achieve both untargeted attacks and targeted attacks. There are more concerned adversarial examples algorithms be proposed in following. In 2016, Papernot et al.\cite{Papernot2017practical} proposed a black-box attack using an alternative model approach. In 2017, Moosavi-Dezfooli et al.\cite{Moosavi2017universal} proposed an universal adversarial example attack, where a particular adversarial interference is able to influence the classification of multiple or even all examples in machine learning. In the same year, Carlini et al.\cite{Carlini2017towards} proposed an optimization-based C\&W algorithm to improve the adversarial examples attack. In 2018, Zhao et al.\cite{Zhao2018generating} found that not need to be filled with artificial interference, but found different distribution in nature and can cause misclassification of machine learning models, which is called Natural Adversarial Examples.

In the real physical world, the environment of autonomous driving is more complex, so there are more challenges for attacker to generate adversarial examples in the physical-world\cite{Kurakin2016adversarial}. The requirements of adversarial examples in the physical world are as follows. 
\begin{itemize}

\item \textbf{Physical generatible.} In the physical world, it is insufficient that only adding perturbations via cyber space. Such perturbations must be capable of being physically generated by printing, 3D printing or spraying, etc.

\item \textbf{Local generatible.} The local nature of the adversarial example. In an  adversarial example of the digital world, the attacker can add perturbations to any pixel within the range of the image; however, in a physical world attack, there are often only local areas of the target that are available, and in many cases, the background areas of the image are difficult to use to generate a physical world adversarial example.

\item \textbf{Robustness.}  In the physical world, especially in the field of autonomous driving, it is often required that the adversarial examples can continuously produce misleading effects on machine learning models during multiple angle changes. At the same time, the adversarial examples need to be continuously effective against mainstream target recognition algorithms under certain distance and angle ranges, multiple natural environments, and multiple resolution sensor devices. This puts forward higher requirements on the persistence and universality of adversarial examples, which greatly increases the complexity of the generation of adversarial examples. 

\end{itemize}
Therefore, more adversarial example robustness enhancement measures are often required to realize the physical world adversarial example attacks; and the specific measures applied to vary according to different scenarios and attack targets. In accordance with the scenarios and objects for which countermeasures are set, this paper divides the recognition targets in the autonomous driving system into three categories: vehicles including various motor vehicles, pedestrians such as walkers and cyclists, and static targets like road facilities, traffic signs, markings, roadside advertising signs and other static objects.

Thus, at present, the physical world in the autonomous driving field universally has 3 types of targets for adversarial example techniques: static objective recognition, pedestrian recognition, and vehicle recognition.

\begin{itemize}

\item Physical world adversarial examples for static targets. Such attack targets include a variety of static target recognition systems such as traffic signs, traffic signals, and traffic markings, which are characterized by the requirement that the adversarial examples can continuously and steadily interfere with the judgments of machine learning models over a large range of angles and distances. In 2017, Lu et al.\cite{Lu2017adversarial} successfully performed adversarial example generation in the physical world for the popular objective recognition algorithm FasterRCNN. To  achieve better distance and angle range adaptation, in 2018, Eykholt et al.\cite{Eykholt2018robust} proposed the Robust Physical Perturbations (RP2) algorithm. In the same year, Chen et al.\cite{Chen2018shapeshifter} adopted the Expectation over Transformation (EoT) method\cite{Athalye2018synthesizing} to improve the generation of adversarial examples for traffic signs, resulting in improved adaptability of the adversarial examples to distance, angle, light and other environments. In 2019, Zhao et al.\cite{Zhao2019object} proposed the feature-interference reinforcement (FIR) algorithm and the realistic constraints generation (ERG) algorithm to enhance the robustness of the adversarial examples. At the same time, they proposed the nested-AE algorithm to improve the adaptability of the adversarial examples to long and short distances. Finally, the composite scheme is able to success attack against popular objective recognition algorithms, such as YOLO v3 and Faster-RCNN, within $\pm 60 ^{\circ}$ angle and $[1m ,25m]$ distance range. Based on the above methods, \textbf{hiding attacks} and \textbf{appearing attacks} can be carried out. The hiding attack is to paste an adversarial example on a normal traffic sign to make the target recognition system fail to recognize the traffic sign. The appearing attack to paste an adversarial example on other objects, causing the target recognition system to recognize the object as a characteristic traffic sign or make a false recognition. In 2020, Kong et al.\cite{Kong2020physgan} proposed PhysGAN, a physical world adversarial examples attack method based on adversarial example generative networks, and the generated adversarial examples of advertising signs have better robustness and invisibility.

\item A physical world adversarial example for pedestrian recognition. In autonomous driving system, a missed detection of pedestrians by the objective recognition system can have serious consequences. In 2020, Wu\cite{Wu2020making} proposed the "invisibility cloak" algorithm, where pedestrians are not normally detected by Yolo v2 and Yolo v3 objective detection models when wearing sprayed adversarial example clothing. In the same year, Wang et al.\cite{Wang2020can} conducted a similar study.

\item Physical world adversarial examples for vehicle recognition. These examples are often pasted or painted on the vehicle body with a specific pattern, so that the vehicle detection system can not identify the vehicle or mistakenly identify the vehicle as other objects. Such physical world adversarial examples try to maintain attack effectiveness under high speed movement, various light and other external conditions, especially in the 360-degree view and within the detection range of vehicle identification system, which puts forward higher requirements for the robustness of adversarial samples. At the same time, at different angles, the camera may only be able to acquire part of the images of he adversarial example in the vehicle body, which in turn places a local requirement on the adversarial example, i.e., part of the adversarial example can also achieve the attack. In 2019, Zhang et al.\cite{Zhang2019camou} proposed a vehicle body painting method of black box adversarial example based on transition models so that the example vehicle could not be identified by autonomous driving vehicle detection system. In 2020, Wu et al.\cite{Wu2020making} proposed the discrete searching algorithm to efficiently generate adversarial patches, and then proposed the Enlarge-and-Repeat (ER) algorithm to extend the adversarial patches to the whole body using body images collected from all angles. Both are with pretty good adversarial results.

\end{itemize}

The following highlights the key algorithms in the physical world adversarial example enhancement described above.


\textbf{Algorithm1. Expectation over Transformation (EoT)} \cite{Athalye2018synthesizing}

The core idea of the EoT algorithm is to add a certain random perturbation to each iteration of the adversarial example generation process, so that the final generated adversarial example has better robustness, with specific transformations including: projecting, rotation, and scaling. In the formula, the operation $M_{t}(x_{b},x_{o})$ is defined to project the target image $x_{o}$ onto the background image $x_{b}$ through some transformation $t$, and then the EoT is optimized as follows.

\begin{equation}
\begin{split}    
 \hat{p} = &\arg\min_{x'\in \mathbb{R}^{h\times w \times 3}} \mathbb{E}_{x \sim X, t\sim T}[L(F(M_{t}(x,tanh(x'))] \\ 
&+ c \cdot \left \| tanh(x') - x \right \|  
\end{split}  
\end{equation}

where $X$ denote the training set of background images, and $F$ denote the target network.


\textbf{Algorithm2. Adversarial Patch }\cite{brown2017adversarial}

Based on the EoT transform, the attacker could generate an adversarial patch $\hat{p}$ which the image with the this adversarial patch attacked by adversarial examples. The adversarial patch can be any shape, transformed by EoT methods such as random projecting, rotation, and scaling, and then generated by optimization methods such as gradient descent. An adversarial patch generation task $A(p,x,l,t)$ can be formally described as: for any particular $x \in \mathbb{R}^{w \times \ h \times c} $  to generate adversarial patch $p$  through the EoT transformation $t$ at position $l$, then  the adversarial path $p$ is continuously optimized by the following optimization algorithm:

\begin{equation}
\hat{p} = \arg \max_ {x\in \mathbb{R}^{h\times w \times c}}  \mathbb{E}_{ x \sim X, t\sim T, l \sim L}[\log Pr( \hat{y}| A(p,x,l,t))]
\end{equation}

where $X$ denotes the training set of background images, $T$ denotes the distribution of the EoT transform used by the patch, and $L$ denotes the distribution of the locations of the adversarial patches.



\textbf{Algorithm3. Feature-inference Reinforcement}\cite{Zhao2019object}

Adversarial example generation algorithms often require an objective function, or called it as loss function, designed to minimize the difference between the predicted and expected values of a deep learning model. The neural network extracts features of the objects in the image and makes classification predictions based on these extracted features. The researchers found that generating adversarial examples with perturbations further forward in the hidden layer in the neural network would make the adversarial examples more robust. The core idea of the FIR algorithm is therefore to minimize the difference between the feature images of the layers of the adversarial examples and the normal examples, except for the use of the adversarial examples to mislead the prediction results of the neural network.

First, attacker obtain the feature images $Q_{n}$ and $Q'_{n}$  generated by each hidden layer of the neural network for corresponding categories $y$ and $y'$. Next, generate the feature vectors $v$ with $v'$ from the feature images $Q_{n}$ and $Q'_{n}$. Finally, attacker optimize with the a composited loss function until convergence, then return the adversarial example. Such a vector loss can be defined as $Loss_{f} = \sum \left | v -v' \right |$. The FIR algorithm can be described formally as follows.



    \begin{algorithm}[h]
	\caption{\textbf{3. Feature-inference Reinforcement}}\label{alg:cap}
    \KwInput{normal example $x$, target objective detection model $f$}
    \KwOutput{adversarial example $x'$}
    \LinesNumbered
    \SetAlgoLined
     
      \While  {until convergence } {
         
         \ForEach {$i$th hidden layer in $f$ }{
               $q_{i} \gets f(x) $ //get feature map $q_{i}$ of normal examples $x$ with prediction $y$ ;\\
              
                $q'_{i} \gets f(x') $ // get feature map $q'_{i}$ of adversarial examples $x'$ with prediction $y'$ ; \\                                                                            }
      $v \gets Q:\left \{ {q_{1},q_{2},...,q_{n}} \right \}$  // generate normal feature vector from feature maps
      
     $v' \gets Q':\left \{ {q'_{1},q'_{2},...,q'_{n}}\right \} $   // generate adversarial feature vector from feature maps
    
    Optimize $x'$ to minimize the loss function:   $$ \alpha C_{N}^{box} + \beta p_{N}(y^{N} | S ) + c (loss_{f})^{-1}   $$  
    }
    
    \end{algorithm}

where $C_{N}^{box}$ denotes the confidence level of the target detection system for the target area; $y^{N}$ denotes the confidence level that example $x$ is judged by the neural network to be class $N$; $S$ denotes the distribution space of confidence levels; $loss_f $ denotes the vector loss; $\alpha$, $\beta$ and $c$ are three constants representing the weights respectively.



\textbf{Algorithm4. Nested adversarial examples (Nested-AE)}\cite{Zhao2019object}

Most objective detectors were designed to use multi-scales, each works better at different distances individually. It means that at different distances, different scales play different roles. In order to make the generated adversarial examples achieve adversarial effects over a large distance range and multiple angles, the Nested-AE algorithm considers many scales for different distances and angles to generate many adversarial patches. Then Nested-AE obtained the adversarial patches to synthesize an adversarial example. Thus nested adversarial examples are more adapted to the environment of vehicle movement in autonomous driving, which can achieve the purpose of a continuous adversary effect on the objective detection system. The nested adversarial examples could be formally described as

\begin{equation}
X^{adv}_{i+1} = Clip 
\begin{Bmatrix}
X_{i} + \varepsilon sign(J(X_{i})), &S_{P} \leq S_{thres}  \\ X_{i}+ \varepsilon M_{center}sign(J(X_{i})) , &S_{P} > S_{thres} 
\end{Bmatrix}  
\end{equation}

where $X_{i}$ denotes the original example, $X^{adv}_{i+1}$ denotes the example with adversarial perturbation, $J()$ represents the gradient of input $X_{i}$ and $Clip()$ refers to regularize the input to the $[0,255]$ interval.

If the adversarial example size $S_{p}$ is less than or equal to the threshold $S_{thres}$ , then it is considered a long-range adversarial attack at that point and the whole will be perturbed ; otherwise it is a close-range attack and only the central region will be perturbed. This decomposes the adversarial attack at different distances into two sub-tasks, which are perturbed and optimized separately.

The objective detection systems usually divides each video frame into a grid that consisted of $m \times n$ boxes. Base on the prediction of each box, attacker could find the decisive box for each scales, and then add the adversarial perturbation in this box. As the predicted output is usually tensors, there are only need the tensor of the index where the adversarial example is located. Therefore, it can calculate the index by the size of the example and the position of the centre region, so that we can get the tensor representing the example region, which is denoted as $N_{p}$. Then $N_{p}$ needs to be calculated in each video frame. The nested adversarial example algorithm can be optimized using the following loss function.

\begin{equation}
N_{p} = f(p_{size},P_{position}),
1-C^{box}_{N_{p}} + \beta \sum \left | p_{N_{p}},j - y_{j} \right |^{2}
\end{equation}


\subsection{Physical World Adversarial Example for Pedestrian Detection}

Pedestrian detection often places high demands on the universality, portability, robustness and feasibility of the adversarial examples. To due this problem, some algorithms was proposed. As a typical example, Invisibility Cloak algorithms\cite{Wu2020making} can generate adversarial examples in the physical world, so that  pedestrians wearing clothes painted with specific adversarial examples cannot be recognized properly. If the similar method be used in pedestrian,  serious consequences of autonomous driving system will be lead. 


\textbf{Algorithm 5. Invisibility Cloak}\cite{Wu2020making}

The policy of the invisibility cloak algorithm is to use a large number of images containing people to train against patches; in each iteration, a random batch of images is selected and sent to the objective detection system to obtain the bounding box of people. The critical idea is that place a randomly transformed patch on each detected person so that the score that feature images are detected to have people’s presence is minimized. 

The patch $P \in \mathbb{R}^{w \times h \times 3} $  is projected to the target image $I$ by transformation function $R_{\theta}$ which performs data augmentation for lighting, contrast changes, distortion, with $\theta$ as the parameter, in addition to scaling the patch to fit the size of image $I$.  

Additionally, two effective advanced methods for physical world attack were proposed as auxiliary. One was called \textbf{Total-Variation (TV) loss.} 
Add a TV penalty function to the patch to make adversarial example smoother to improve robustness of physical world adversarial examples.

with the final loss function as: 

\begin{equation}
      L_{obj}(P) =  loss + \gamma \cdot TV(P)  
\end{equation}

Another is \textbf{Ensemble training}. In the black-box case, the attacker cannot obtain the gradient of the target detection model under attack, so a possible solution is to collect multiple similar white-box models for ensemble training. The loss function is 

\begin{equation}
      L_{ens}(P) = \mathbb{\theta,I} \sum_{i,j} \left \{ S_{i}^{(j)}(R_{\theta}(I,P)) + 1,0 \right \}  
\end{equation}
     
where $S_{(j)}$ represents the objective detection model of target $j$. It is formally described as follows.

\begin{algorithm}[h]
\caption{\textbf{5. Invisibility Cloak}}\label{alg:cap}
\KwInput{normal example $x$, target objective detection model $f$}
\KwOutput{adversarial example $x'$}
\LinesNumbered
\SetAlgoLined
\setcounter{AlgoLine}{0}

$I = P(x)  $                   // generate projection   \\  \;

$  R_{\theta} \gets \theta  $     // define a transfer function  $R_{\theta}$       \\  \;

\While  {until convergence } {

Optimize $x'$ to minimize the loss function: 
$$ L_{obj}(P) =  R_{\theta}(I,P) $$
                                                 }
                                                 
or Auxiliary1: TV Loss
$$ L_{obj}(P) =  R_{\theta}(I,P) + \gamma \cdot TV(P) $$

or Auxiliary2: Ensemble training
$$ L_{obj}(P) = \mathbb{E}_{\theta,I} \sum_{i} \max \left \{  {S_{i}(R_{\theta}(I,P))+1,0} \right \}^{2} $$

\end{algorithm}


\subsection{Physical World Adversarial Example for Vehicle Detection}

The aim of the adversarial examples for vehicle detection is evade or mislead other vehicle detector by adversarial spraying. Different from other scenarios,  the challenge is that the spraying need works for the detector at all angles. Discrete Searching is a typical and effect algorithm.


\textbf{Algorithm 6. Discrete Searching}\cite{Wu2020physical}

The discrete searching algorithm is essentially a black-box adversarial example generation algorithm based on a genetic algorithm that continuously optimizes the adversarial examples through mutation and selection. The discrete searching algorithm that mutation-based search method defines two mutation strategies. One is random mutation, where a point within the circle with $\epsilon$ as the radius, is randomly selected as the direction and advanced a step length as the new mutation point. If the current mutation optimization fails to outperform the original one, the choice will be made to continue with the random mutation. The other is directed mutation, in which the candidate's best mutation point is selected within a particular angular expansion of the current direction, advancing a random step length. If the current variation outperforms the original one, the choice will be made to continue with the directed mutation.

\begin{algorithm}[h] 

\caption{\textbf{6. Discrete Searching}}\label{alg:cap}

\KwInput{normal example $x$, target objective detection model $f$}
\KwOutput{adversarial example $x'$}
\LinesNumbered
\SetAlgoLined
\setcounter{AlgoLine}{0}

\While  {until convergence } {
\ForEach {$i \in \left\{ 0, ..., N_{a}-1 \right\} $ }{
    
     $C_{i}^{j_{w}}$, $C_{i}^{j_{w}} = Clip (C_{i}+Random(H,W,3) \cdot \epsilon_{1} \cdot \delta )$  //generate candidate point
     
     select $\hat{C}_{i}$ in $C_{i}^{j_{w}}$
     
     \eIf{$\hat{C}_{i}$ is better than $C_{i}$}{
     $C_{i}=\hat{C}_{i}$
     }{
     $C_{i}$ remains unchanged
     }
}
}
\end{algorithm} 


\subsection{Adversarial Examples on LiDAR and RaDAR }

Adversarial examples are related to the characteristics of deep learning itself, and both RGB image-based and LiDAR- or RaDAR-based objective recognition systems are likely to suffer from adversarial examples attacks. In 2019, Cao et al.\cite{Cao2019adversarial} proposed an adversarial examples attack method for LiDAR target recognition. The attacker emits a small amount of perturbation laser at the LiDAR system, which led to a small perturbation in the imaging of the LiDAR system, and such perturbation made the LiDAR-based 3D objective recognition erroneous. Then, an adversarial object attack method LiDAR-Adv was proposed for LiDAR objective recognition, in which the attacker could construct a certain special shape of objects, causing the LiDAR objective recognition system to mispredict special objects. Such an attack can be targeted, i.e., the real object is recognized as the specified by the attacker, so this is more easier to exploit for the attacker. As the figure shows, some specially shaped objects can be misidentified as "pedestrians" by the 3D objective recognition system, while others are not recognized properly by the 3D objective recognition system, posing a security risk. The above method was successfully tested on Baidu's autonomous driving system Apollo. In 2020, after research and improvement, SUN et al.\cite{Sun2022counteracting} implemented a black-box attack of the above method, which was successfully tested on Intel’s autonomous driving simulation system Carla. In real autonomous driving environments, the detection of dynamic targets often takes a multi-modal fusion of RGB images, LiDAR, and RaDAR for target recognition, which improves the robustness of the system to some extent.

\begin{figure}[htb]
\centering
\includegraphics[width=\linewidth]{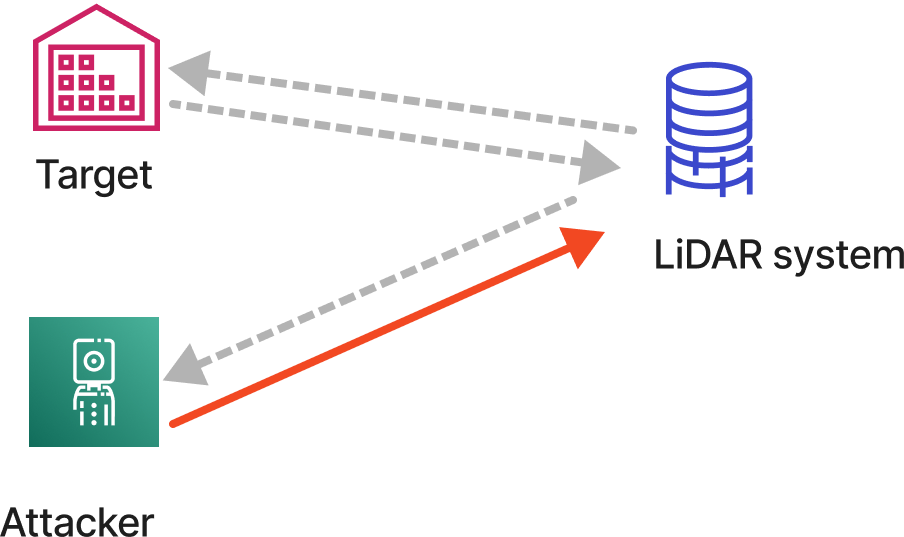} 
\caption{Adversarial Examples on LiDAR\cite{Cao2019adversarial}}\label{Fig3.1}
\end{figure}

Higher demands are placed on the adversarial example for multi-modal environments, mainly in terms of:

\begin{itemize}

\item Adversarial perturbation needs to be able to be physically generated in both the RGB image and the LiDAR system environment. Traditional RGB adversarial examples usually change the RGB values of some pixels in the image, however, this method cannot works on the 3D cloud point map generated from the LiDAR. On another hand, emit a specific adversarial laser at LiDAR can interfere with the LiDAR system, but it is also difficult to effectively influence the RGB objective recognition system.

\item The adversarial examples need to be able to physically and continuously work on both the RGB-based system and the LiDAR-based system. In a real vehicle environment, the RGB system and the LiDAR system need to attack successfully and continuously at a long distance and at different angles.

\item Adversarial examples need to be able to adapt to different data preprocessing between RGB-based systems and LiDAR-based systems. RGB image acquisition system and LiDAR data acquisition system both have certain data preprocessing, which will have impact on adversarial examples. The algorithm of adversarial examples generation needs to have strong robustness to different preprocessing.

\end{itemize}
Meanwhile, after optimized by some specific algorithms , the threat to the fused  objective recognition system by adversarial examples is still exists. In 2021, Cao et al.\cite{Cao2021invisible} proposed the MSF-ADV method with LiDAR and RGB image fusion environment as an example, and successfully implemented the physical world adversarial examples.


\textbf{Algorithm 7.  MSF-ADV Algorithm}\cite{Cao2021invisible} 

To accommodate the above challenges, MSF-ADV first generates 3D objects of different shapes so that they can simultaneously affect the LiDAR-based 3D point cloud imaging and also the RGB colour values of the pixels in the RGB image. Secondly, MSF-ADV uses an optimization algorithm to generate the 3D shapes with the best adversarial effect. Finally, MSF-ADV uses a 3D printer for physical generation. The loss function of the optimization algorithm can be described as

\begin{equation}
 min_{S^{a}} \mathbb{E}_{t \sim T } [\mathcal{L}_{a}(t(S_{a});\mathcal{R}^{l},\mathcal{R}^{c},\mathcal{P},\mathcal{M}) + \lambda \cdot \mathcal{L}_{r}(S^{a},S)]  
\end{equation}

where $S$ denotes the original examples the $S_{a}$ denotes the adversarial examples, $\mathcal{M}$s denotes the fusion algorithm, and $\mathcal{R}^{c}$ is the derivative projection function used to represent RGB image based prediction. $\mathcal{R}^{l}$ is the derivative projection function used to represent the LiDAR prediction. $\mathcal{P}$ is the ultimate output of the objective recognition system.

Through thus optimization, the trade-off between RGB colour and LiDAR shape was found.


\subsection{Adversarial Examples on Object Tracking \& Trajectory Prediction}

Usually, autonomous driving system relies on object tracking and trajectory prediction, to determine and predict target states, and to support driving decisions. Object trajectory tracking can be divided into Single-Object Tracking (SOT) and Multi-Object Tracking (MOT). With the application of object tracking in critical cyber systems, adversarial examples attacks on it are also rising. Among them, the main purpose of the adversarial examples attack on SOT is to achieve objective evasion. In 2020, Chen et al.\cite{Chen2020one} proposed the one-shot adversarial attack, which only adds a weak perturbation to the initial frame in the video, and the tracked object may not be able to track the trajectory in subsequent frames. In the same year, Yan et al.\cite{Yan2020cooling} proposed the cooling-shrinking attack, which perturbs the object search area by adding specific adversarial noises, so that the tracker cannot identify the object and interrupted the trajectory tracking. In the next year, Jia et al.\cite{Jia2021iou} proposed the IoU Attack, the idea of which is to reduce the fractional difference between the normal object border and the adversarial object border in object tracking, thus enabling the trajectory offset using SOT system.

The autonomous driving system more often uses MOT systems. An adversarial example attack on MOT may achieve both evasion and object obfuscation. In 2020, Jia et al.\cite{Jia2020fooling} generated an adversarial example on an autonomous driving object tracking system that minutely deviation the normal target identification bounding box of the attacked target in a specific direction, causing the tracker to assign the wrong velocity to the attacked trajectory, resulting in the target tracking system not being able to associate with the target properly, thus achieving an escape attack. In 2021, Lin et al.\cite{Lin2021trasw} proposed a new adversarial example scheme that mainly uses the "PullPush Loss" algorithm and "Center Leaping" algorithm. The scheme leads to object tracking system confuses, when objects cross each other. 


\textbf{Algorithm 8. Push-Pull Loss} \cite{Lin2021trasw}

A video $V$ consists of a series of frames, which can be marked as $V=\left \{{I_{1},I_{2},...,I_{N}}\right \}$, The trajectories of target$i$ and $j$ are respectively $T_{i}=\left \{{O_{s_{i}}^{i}},...,O_{t}^{i},...,O_{E_{i}^{i}}\right \}$ and $T_{j}=\left \{{O_{s_{j}}^{j}},...,O_{t}^{j},...,O_{E_{j}^{j}}\right \}  $

The attacker's target generates a series of adversarial frames $\hat{V}$, as $\hat{V} =\left \{ {I_{1},...,I_{t-1},\hat{I}_{t},...,\hat{I}_{t+n-1},I_{t+n},...,I_{N}}\right \}$, such that from the moment of time $t$,
an adversarial misdirection of the trajectory of $i$ and $j$ occurs. Then there is the formula, $\hat{T}_{i}=\left \{{O_{s_{i}}^{i},...,O_{t-1}^{i},O_{t}^{i},...,O_{t+n-1}^{j},O_{t+n}^{j},...,O_{e_{j}}^{j}}\right \}$ where $O_{t}^{i}$ indicates the target identified as $i$ at the moment of time $t$.

Use the PushPull loss function to optimize and realize:

\begin{equation}
\begin{split}
L_{pullpush}(a_{t-1}^{i},a_{t-1}^{j},feat_{t}^{i},feat_{t}^{j})  \\
= \sum_{k \in \left\{i,j \right\}} d_{feat}(a_{t-1}^{k},feat_{t}^{\widetilde{k}}) - d_{feat}(a_{t-1}^{k},feat_{t}^{k}) 
\end{split}
\end{equation}

where $d_{feat}()$ denotes the cosine distance and $a_{t-1}^{i}$ and $a_{t-1}^{j}$ represent the trajectory features of object $i$ and $j$ while  $feat_{t}^{i}$ and denotes the features of object $i$ and object $j$. After continuous optimization, it make the adversarial feature $feat_{t}^{j}$ instead of $feat_{t}^{j}$ be classified as trajectory $k$.


\begin{figure}[htb]
\centering
\includegraphics[width=\linewidth]{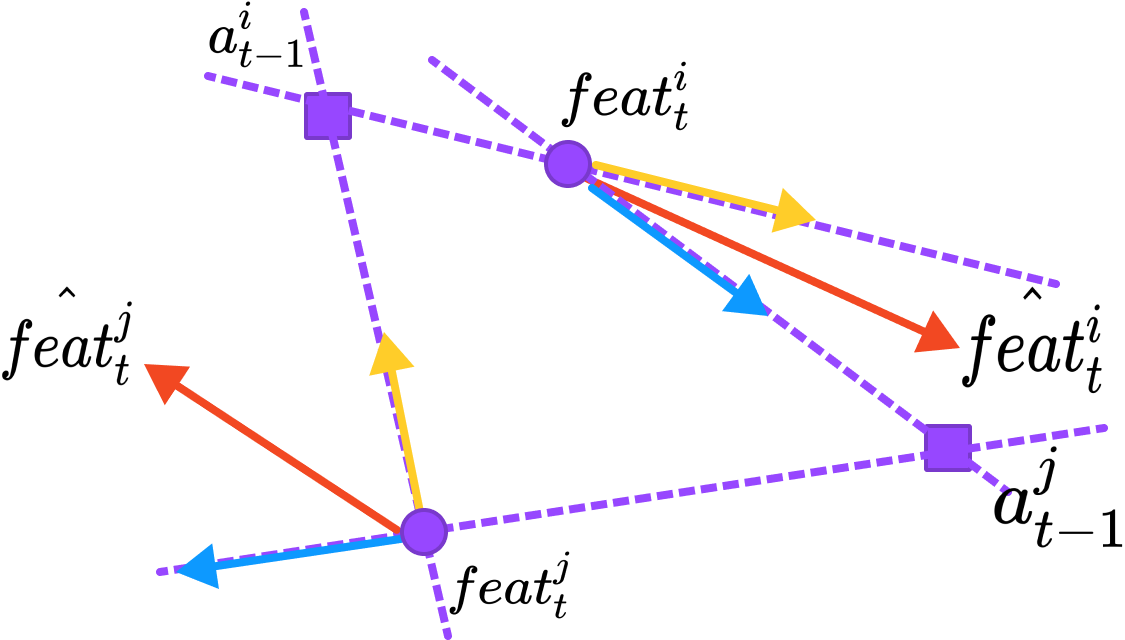} 
\caption{Push-pull Loss Function \cite{Lin2021trasw}}\label{Fig3.7}
\end{figure}


\textbf{Algorithm 9.  Center Leaping} \cite{Lin2021trasw}

With the PushPull loss function optimization described above, it is able to make the object tracking misled against attacks, which is still difficult to succeed when the difference between the two objects is large. The idea of the Center Leaping algorithm is to first mislead the objective recognition link so that the objective candidate box identified by the target recognition system is shifted towards the target to be misled, thus achieving a better attack success rate when there is a large deviation in the distance and size difference between the two tracked objects. The loss function is

\begin{equation}
\begin{split}
L &= min \sum_{ k \in \left \{ i,j \right \}} d_{box}(K(m_{t-1}^{\widetilde{k}},box_{t}^{k}))  \\
&=  min\sum_{ k \in \left \{ i,j \right \}} d(cent(K(m_{t-1}^{\widetilde{k}},box_{t}^{k}))  \\
&+  min \sum_{ k \in \left \{ i,j \right \}} d(size(K (m_{t-1}^{\widetilde{k}},box_{t}^{k}))   \\
 &+min \sum_{ k \in \left \{ i,j \right \}} d(off(K(m_{t-1}^{\widetilde{k}},box_{t}^{k}))) 
\end{split}
\end{equation}

Of which $m_{t}^{k}$ and $box_{t}^{k}$ respectively represent the trajectory state and the candidate frame of target $k$ at time $t$; $cent()$, $size()$, and $off()$ respectively represent the centre point position, the size, and the offset of the candidate box; and $d()$ represents distance $L_{1}$.

The central leaping algorithm can be expressed as

\begin{equation}
\begin{split}
L_{cl} = \sum_{ k \in \left\{ i,j \right\} }  (& \sum_{(x,y) \in B_{c -> \widetilde{k}}}(1-M_{x,y}^{\gamma} log(M_{x,y}) + \\
&\sum_{(x,y) \in B_{c->k}} (M_{x,y}^{\gamma}log(1-M_{P}{x,y})))
\end{split}
\end{equation}

$M(x,y)$ denotes the heat value of $(x,y)$, $c_{k}$ denotes the center of the object, $c \rightarrow  \widetilde{k} $ denotes the direction from $c_{k}$ to $cent(K(m^{\widetilde{k}}_{t-1}))$. During the optimization process, the center point will move to the adjacent grid along this direction. In the object recognition and tracking system, the heat value of the original target center will drop, and the heat value in the direction close to the object will rise, so as to achieve the goal of the candidate frame approaching the object.

Similarly, it is able to integrate the size loss function and the offset loss function into a A novel composite loss function.

\begin{equation}
\begin{split}
L_{reg} &= L_{size} + L_{off} \\
&= \sum_{k \in \left\{i,j\right\}} L_{1}^{smooth}(size(K(m_{t-1}^{\widetilde{k}}),size(box_{t}^{k}))  \\
&+ \sum_{k \in \left\{i,j\right\}} L_{1}^{smooth}(off(K(m_{t-1}^{\widetilde{k}}),off(box_{t}^{k})) 
\end{split}
\end{equation}

where $L_{1}^{smooth}$ is smooth loss function based on $L_{1}$:

\begin{equation}
 L_{1}^{smooth}(a,b) = \left\{\begin{matrix}
0.5 \cdot (a-b)^{2} & if |a-b| <1 \\ 
|a-b|-0.5 & else 
\end{matrix}\right.  
\end{equation}

\begin{figure}[htb]
\centering
\includegraphics[width=\linewidth]{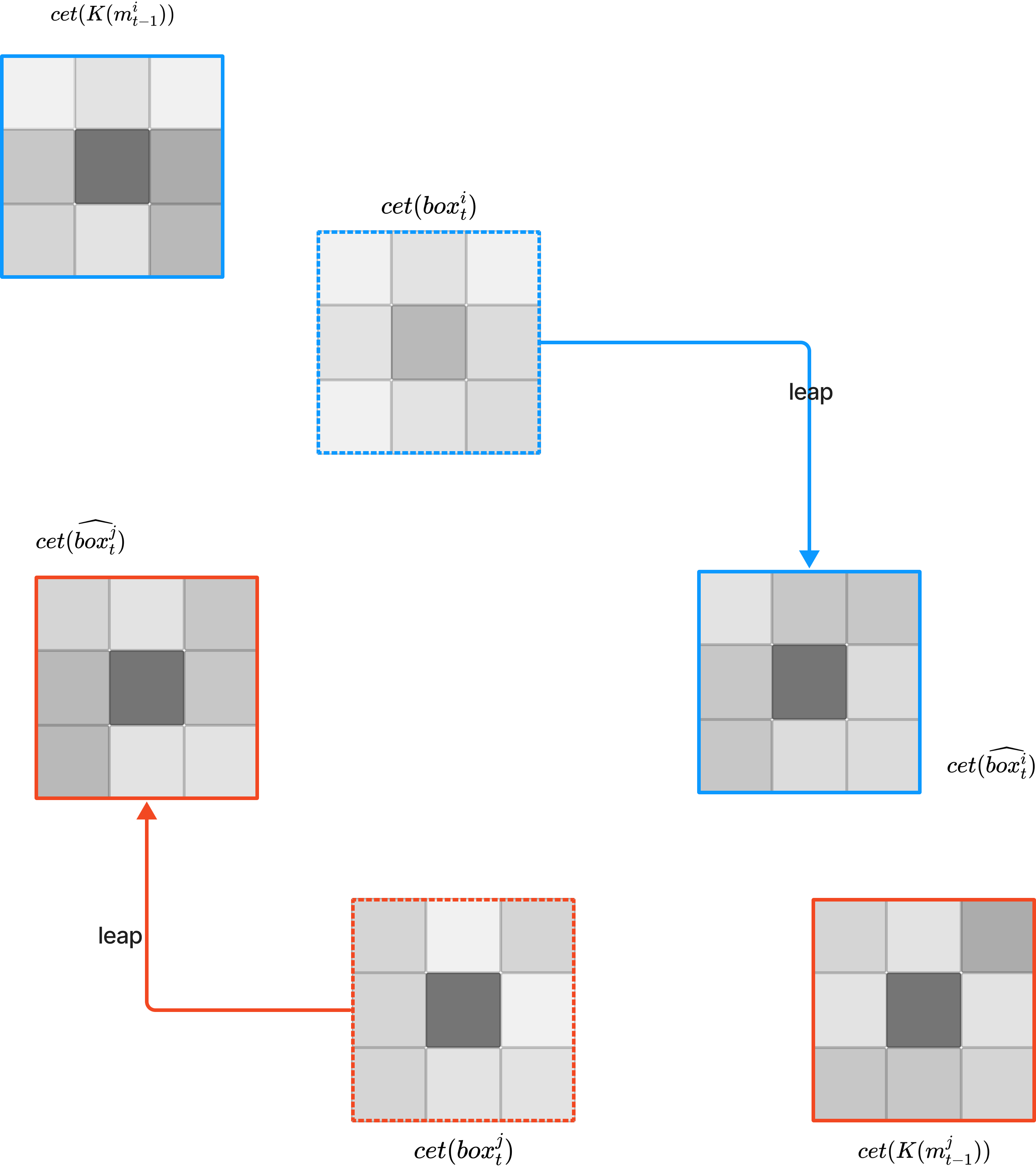} 
\caption{Centre Leaping Principle\cite{Lin2021trasw} }\label{Fig3.8}
\end{figure}

\subsection{ AI Backdoor \& Poisoning on ADS}

Artificial intelligence models are often generated from a certain amount of training data. Some scholars have found that if the training data is not trustworthy, it may lead to the generation of models with "backdoor", which can be manipulated by attackers in the subsequent use of the models, causing serious security risks. Currently, the concepts of "AI backdoor", "AI model poisoning" and "AI Trojan horse" have some similarities, but are not expressed in the same way in different literature. One type of attack is called Training-only attacks, or \textbf{Poisoning Attacks} are usually defined as attacker contaminating part of the training data or modifying the labels of the training data. On the contrary, another type of attack is called \textbf{Backdoor Attacks} or AI Trojans\cite{Goldblum2020data}, in which attackers must participate in both training and testing.  Both poisoning attacks and AI backdoor attacks can cause serious security threats to autonomous driving system, and this paper refers to these two types of attacks as "AI backdoor attacks", where a specific malicious modification is made to a target model in a specific way, causing the model to make harmful judgments about a specifically predicted example. There are similarities and differences between backdoor attacks and adversarial example attacks. Adversarial examples usually do not change the model itself and will not damage the integrity of the AI model, but mainly interfere with the test examples and affect the availability and correctness of the machine learning model. On another hand, AI backdoor takes the form of modification of the AI model, poison of the training data, aggregation of the backdoor model, etc., causing tiny changes to the AI model, affecting both the integrity of the AI model. AI backdoor attacks tend to be more hidden, highly universal, and more damaging. So far, there are two major methods to implement AI backdoor, one is Data Poisoning Attack, and the other is Model Poisoning Attack. Data Poisoning Attack means attacker adds a small amount of poisoning data into training dataset, so that the resulting AI model has a backdoor, and the AI model may make a specific judgement when the predicted example contains a "trigger". Model Poisoning Attack means the attacker directly modifies the model or indirectly fuses the target model with a harmful model by using model integration, federated learning, and transition learning, causing the model to make a directed and erroneous judgment on a specific prediction example.

It has been argued that AI backdoor attacks already exist in traditional machine learning. In 2008, Nelson et al. \cite{Nelson2008exploiting,Barreno2010security} proposed the backdoor attack on Bayesian networks. In 2012 Biggio et al.\cite{Biggio2012poisoning} proposed backdoor attack on SVMs. In 2016, Alfeld et al.\cite{Alfeld2016data} proposed backdoor on auto-regressive prediction models. In 2017, Gu et al.\cite{Gu2017badnets} first proposed backdoor attack on deep learning, then AI backdoor became a promising research topic. The BadNet algorithm adds a small number of training data with the pre-designed pattern into the training data and labels such training examples with a specific target, then the trained model is likely to predict examples with "Trigger" according to the attacker. In the same year, Muñoz-González et al.\cite{Munoz2017towards} proposed a gradient-based algorithm for AI data poisoning. However, for  autonomous driving system, the basic AI backdoor algorithms described above have two  limitations.
\begin{itemize}

\item Control right of training data by attacker.As it requires the attacker to be able to contaminate a certain amount of training data, this requires the attacker to have some control over the training data; at the very least the attacker needs to have background knowledge of the target model's structure, parameters, etc., which places certain requirements on the attacker.

\item Concealment of attack. It relies on contaminating part of the training data by adding a 'pattern' or changing the label of the data. Although the pattern may be relatively insidious, forcing patterns into normal examples may cause a certain amount of unnaturalness that may be detected by humans, or possibly by automated detection through some anomaly identification method. It may also lead to human feel a sense of inharmonious if the attacker modifies excessive label of the training data.

\end{itemize}
In response to these limitations, researchers have made a number of subsequent improvements. On the one hand, attackers have improved the concealment of poisoning attacks by enhancing the concealment of the patterns in the example or minimizing the impact on the integrity of the label. One of the research directions is "clean label", which aims to keep the label of poisoned example semantically correct while realizing data poisoning. In 2018, Shafahi et al.\cite{Shafahi2018poison} proposed the Poison Frogs algorithm, which was the first to implement the Clean Label attack for deep learning. In the same year, Truner et al.\cite{Turner2018clean} proposed two methods of data generation based on adversarial network and adversarial example to achieve a label-consistent "clean example" attack. Another research direction is "Hidden Trigger", also known as "Invisible Trigger", which aims to optimize the trigger pattern to make it as invisible as possible to escape detection by humans and machines.  In 2018, Suciu et al.\cite{Suciu2018does} , and in 2019 Saha et al.\cite{Saha2020hidden} put forward "hidden trigger" which can generate trigger patterns that humans are unable to directly perceive through the senses. In 2020, Wallace et al.\cite{Wallace2019universal} devised a "hidden trigger" poisoning attack in the field of natural language processing. In the same year, Li et al.\cite{Li2020invisible} used information hiding and regularization methods to  improve the bad net algorithm to improve the invisibility of the trigger pattern. 

On the other hand, the attacker reduces the proportion of contaminated training data as much as possible, or even implements a black-box attack that does not require contaminating data, thus reducing the background knowledge required for the attack and lowering the threshold for implementing the attack. In 2017, Liu et al. \cite{Liu2016delving} implemented a black-box approach to generate backdoor by exploiting the migratory nature of the attack, but such backdoor mainly exists in the fully connected layer at the end of the AI model, which can easily fail once the model is fine-tuned; in the same year, Chen et al.\cite{Chen2017multi} proposed a machine learning-based approach to generate AI backdoor, which eliminates the need for attackers to understand the structure of the target system and other information, and reduces the background knowledge requirement; in 2019, Yao et al.\cite{Yao2019latent} proposed "latent triggers", which are first generated in the "teacher model" and then migrated to the "student model" through transition learning. The backdoor is not only found in the last fully connected layer of the student model, but also in all of its layers and thus the difficulty of detecting the "backdoor" through analysis is increased. In the same year, Zhu et al.\cite{Zhu2019transferable} investigated the migratory nature of clean label attacks and used knowledge transition to realize a black-box clean label attack.

In the field of autonomous driving, in 2018, Liu et al.\cite{Liu2017trojaning} realized AI backdoor attacks in a variety of environments, including simulated autonomous driving platforms. In 2019, Rehman et al.\cite{Rehman2019backdoor} implemented an AI backdoor attack on traffic signs in the physical world; Barni et al.\cite{Barni2019new} conducted a clean label poisoning attack on traffic signs; Ding et al. from Nanjing University\cite{Ding2019trojan} designed a "natural trigger" for autonomous driving system to trigger AI model backdoor in special weather like a rainy day, to make red lights incorrectly identified as green lights and numbers incorrectly identified in a specific way; Yao et al.\cite{Yao2019latent} from the University of Chicago used their proposed "latent trigger" method to generate backdoor traffic signs for a variety of models, generating human-imperceptible triggers on physical traffic signs. In 2021, Tian et al.\cite{Tian2021poisoning} achieved a clean label attack on 3D cloud point map. In 2022, Udeshi et al.\cite{Udeshi2022model} proposed an anti-backdoor attack method that can be used in traffic sign recognition scenarios, which achieved avoiding AI backdoor attacks by filtering the triggers in the captured images and correcting the prediction examples.


\textbf{Algorithm 9.  Feature Collisions}

Feature collision is the more common method of AI backdoor generation, where the attacker first selects a target instance from the test set. To achieve poisoning, the attacker chooses a base class instance from the base class and makes imperceptible changes to it, thereby generating a poisoned instance that is injected into the training data later; then, during the training phase, the model is trained using a poisoned data set consisting of a clean data set plus poisoned instance; in the reasoning phase, this causes the target instance to be mistaken by the misclassification model for being in the base class during testing. It is described formally as follows.

$f(x)$ represents the neural network's prediction on input example $x$. Example $x$ colliding with the target is found in the feature space and then computed to be close to the base instance $b$. The target function is

\begin{equation}
p = arg \min \left \| f(x) - f(t) \right \|^{2}_{2} + \beta \left \| x -b \right \|^{2}_{2} 
\end{equation}

$p$ is the poisoning instance, which will be misdirected as the attack target.


\subsection{Summary}

The risk of adversarial example and AI backdoor is brought by the characteristics of deep learning itself. Whether an autonomous driving system uses RGB cameras, LiDAR, RaDAR or other sensors as the source of information collection, it often dependent on deep learning for perception and driving decisions. Going with it, there are new safety risks associated with artificial intelligence. At the same time, the autonomous driving system is a huge system, and in the perception layer alone, they consist of many links that rely on deep learning technologies, such as target recognition, image segmentation, depth estimation and target tracking, which constitute a complex decision-making process, and each link is also subject to different types of AI security threats. It is necessary to ensure the safety of each link to constitute a set of safe autonomous driving system.

Generated adversarial examples are hard to consistently fool neural network classifiers in the physical world. In this chapter, we introduced emphatically method of physical world adversarial examples enhancement, summarized it in Table 2.

\begin{table*}[!h]
\caption{Physical World Adversarial Examples Enhance Methods}
\label{table7.1}
\begin{center}
\begin{tabular}{ p{4cm}<{\centering} | p{8cm}<{\centering} | p{4cm}<{\centering}  }
 \hline
 \hline
 \rowcolor{lightgray}
 \textbf{Method} & \textbf{Contributions} & \textbf{Scenarios in Autonomous Driving} \\ 
 \hline
 \hline
 
 \rowcolor{green!50!yellow!10}
 EoT\cite{Athalye2018synthesizing},  & 
 EoT generate adversarial examples over a chosen distribution of transformations.EoT is the first algorithm that produces robust adversarial examples, which single adversarial examples to an entire distribution of transformations & 
 Object detection  \\
 
 Adversarial Patch\cite{brown2017adversarial} & 
 This attack generates an image-independent patch that can then be placed anywhere within the field of view of the classifier, and causes the classifier to output a targeted class. & 
 Object detection  \\
 
 \rowcolor{green!50!yellow!10}
 FIR\cite{zhao2019seeing} & 
 FIR generated adversarial examples to impact both hidden layers and the final layer. Therefore, the misclassification for adversarial examples depends more on the prior layers in the neural-network, which lead to be more robust in physical scenarios. & 
 Traffic sign detection \\
 
 Nested-AE\cite{zhao2019seeing} & 
 Nested-AE contains two or more Adversarial examples inside that for different distances or angles. It significantly improve the robustness of adversarial attack at the various position. & 
 Traffic sign detection \\
  
 \rowcolor{green!50!yellow!10}
 Randomly Transformed Patch\cite{Wu2020making} & 
 These transforms are a composition of brightness, contrast, rotation, translation, and sheering transforms that help make patches robust to variations caused by lighting and viewing angle that occur in the real world. & 
 Pedestrian detection \\
 
 TV Loss\cite{sharif2016accessorize,komkov2021advhat,Wu2020making} & 
  TV Loss ensures a more smooth patch in which all pixels in the patch get optimized.  & 
 Object detection \\
  
 \rowcolor{green!50!yellow!10}
 Ensemble training\cite{Wu2020making} & 
 Ensemble training fool an ensemble of detectors that were not used for training. & 
 Object detection \\ 
  
 UPC \cite{huang2020universal} &
 UPC optimization constraint to make generated patterns look natural to human observers. &
 Object detection \\
 
  \rowcolor{green!50!yellow!10}
 NPS \cite{sharif2016accessorize} &
 NPS deal with the difference in digital RGB-values and the ability of real printers to reproduce these values. &
 Object detection \\
 
 Discrete Search\cite{Wu2020physical} &
 Discrete Search improve the black-box attack by iteratively refining the camouflage using a mutation-based search method. &
 physical-world Black-box attack \\
 
 \rowcolor{green!50!yellow!10}
 MSF-Adv\cite{Cao2021invisible} & 
 MSF-Adv generate adversarial examples in Lidar, RaDar, and fusion. & 
  Vehicle detection\\
  
Spatial Transformer Layer (STL) to project\cite{komkov2021advhat} &  
 Many kinds of Projects imitate the form changes for rectangula adversarial patches after placing it in physical world. &
Object detection \\

 \rowcolor{green!50!yellow!10}
Sticker Projection\cite{komkov2021advhat} &
Project the obtained adversarial examples with small perturbations in the projection parameters to make the attack more robust. &
Object detection \\
  
 \hline
 \hline
 
\end{tabular} 
\end{center}
\end{table*}

As the foundation, some major general adversarial examples algorithms list in Appendix A.

\section{Emerging Threats of Decision-Making Layer}

The major function of the decision layer is to make the correct driving decision based on sensing and perception. In common autonomous driving architectures, the trajectory of dynamic objects, such as vehicles or pedestrians, must predicted. If the prediction process is maliciously interfered with by an attacker, vehicle may under security threat.

\subsection{Emerging Threat on Prediction-oriented attack techniques}

In general, autonomous driving systems need to predict the short-term or long-term spatial coordinates of various road agents such as cars, buses, pedestrians, rickshaws, and animals, etc. Predicting usually base on Recurrent Neural Network (RNN) techniques, the algorithms of which include LSTM, and Sequence to Sequence. Researchers have proposed attack methods for recurrent neural network algorithms. In 2016, Papernot et al.\cite{Papernot2017practical} proposed an RNN-oriented adversarial example attack, and many subsequent researchers have continued to improve the attack method and enhance the attack effect\cite{Hu2018black,Chang2020audio,Cheng2020seq2sick,Carlini2018audio,Kreuk2018fooling}.

\subsection{Emerging Threat in Imitation Learning}

Imitation Learning and Reinforcement Learning are the two main approaches to driving decision making. Imitation Learning is a data-driven approach that imitates expert driver policies to make decisions\cite{Song2021} and some end-to-end autonomous driving systems use an imitation learning framework\cite{Pan2017agile}. Reinforcement learning, on the other hand, uses deep reinforcement learning algorithms to optimize the model and make the best decisions. Whether a autonomous driving system adopts an imitation learning or a reinforcement learning architecture, an attacker could interfere with the AI model, thereby affecting normal driving decisions to pose a risk to the autonomous driving system.

Imitation learning can be described as a process\cite{Hussein2017imitation}, and human expert experience can be described as a tuple such as $(s,a,r,{s}')$, where s is the state of driving, $a$ is the behavior of the human expert, $r$ is the reward created by the behaviour $a$, and $s'$ is the resulting new state. Imitation learning generates policy $\pi$  through machine learning, based on the captured set of behaviour of the human experts $D=(x_{i},y_{i})$.  

$$ u(t) = \pi(x(t),t,\alpha)$$

$u$ is the predicted behavior given by the machine, $x$ is the feature vector of the state of the environment $s$, $t$ is the time, and $\alpha$  is the set of parameters for a set of policies.

Imitation learning is still a branch of deep learning and based on deep neural networks (DNNs), also it is equally threatened by adversarial examples, AI backdoor and other forms of attack. In the autonomous driving field in 2020, Boloor et al.\cite{Boloor2020attacking} proposed an adversarial example generation algorithm based on Bayesian optimization that can attack end-to-end (E2E) autonomous driving trajectory prediction system, which was successfully experimented on Intel’s Carla simulation platform. This approach sprayed special adversarial patterns generated with the algorithm on the road, interfering with the autonomous driving system's prediction of its own vehicle, to induce the autonomous driving system to make a wrong driving decision. In the same year, Yang et al.\cite{Yang2020finding} proposed two adversarial attack algorithms for vehicle trajectory prediction, which improved the above method, reducing the number of optimization rounds required for adversarial example generation and improving efficiency.


\textbf{Algorithm 10. Bayesian Optimization (BO) algorithm}

The objective of the BO algorithm is to generate adversarial examples suitable for end-to-end autonomous driving system and find the best adversarial perturbation through optimization $\delta$, with an loss function of

\begin{equation}
\delta^{*} = \arg \max_{\delta} f(\delta) 
\end{equation}

where $\delta^{*} \in \mathcal{R}^{d}$, assuming that the autonomous driving model $f$ prediction conforms to a Gaussian process, it can be written as $GP(f,\mu (\delta), k(\delta,\delta'))$, and let the mathematical expectation be 0, then $\mu(\delta)=0$, with the variance is the Mattern covariance function $K$.

\begin{equation}
k(\delta, {\delta}') = (1+ \frac{\sqrt{5}r}{l} + \frac{5r^{2}}{3 l^{2}}) exp(- \frac{\sqrt{5}r}{l})   
\end{equation}

where $r$ denotes the Euclidean distance and $l$ is a factor coefficient. In such a manner, we can consider $\mathcal{l}$ is the adversarial perturbation.


\begin{figure}[htb]
\centering
\includegraphics[width=\linewidth]{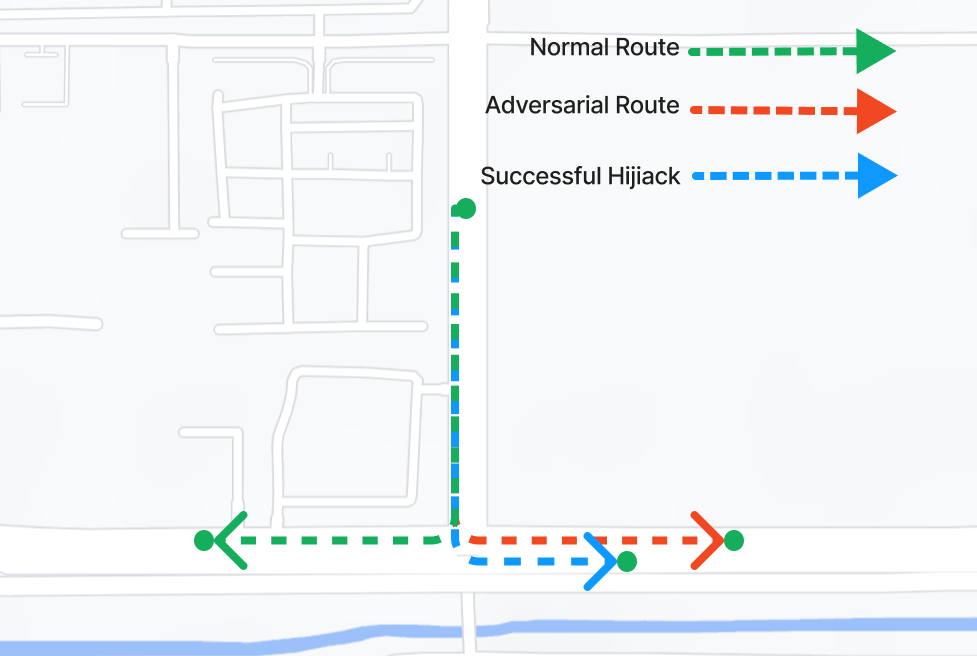} 
\caption{ BO Algorithm \cite{Boloor2020attacking}}\label{Fig3.8}
\end{figure}

\subsection{Emerging Threat on Reinforcement Learning}

Reinforcement learning (deep reinforcement learning) is widely used in fields such as autonomous decision-making, electronic combat and competition. Combined with a number of other techniques such as deep search trees, deep reinforcement learning has enabled \textbf{AlphaGo} to explore some of the blind spots of human cognition. Adopting reinforcement learning technique for autonomous driving decisions is one of the major technology routes in academia and industry. Reinforcement learning security has recently received extensive attention and research.

Reinforcement learning can be described as a Markov Decision Process(MDP)\cite{Arulkumaran2017deep,Hussein2017imitation}. A finite state decision making process consists of the tuple$(S,A,T,R)$ where $S$ is the set of finite states, $A$ is the possible behaviour of the system, and $T$ is the State Transition Probabilities consisting of a set of probabilities $P_{s,a}$. It indicates that when behaviour $a$ is taken, the probability of reaching state $s$ is achieved, and reward function $R$ can return the reward value $Y$ which can be obtained by the reward policy $R(s_{k},a_{k},s_{k+1})$ where the reward value $Y$ denotes changing the state to state $s_{k+1}$ when taking behaviour $a_{k}$ at the state $s_{k}$. Reinforcement learning is the process of starting with a random policy, receiving a reward based on the execution of that policy, and then continuously optimizing the policy by maximizing the reward.

\begin{figure}[htb]
\centering
\includegraphics[width=\linewidth]{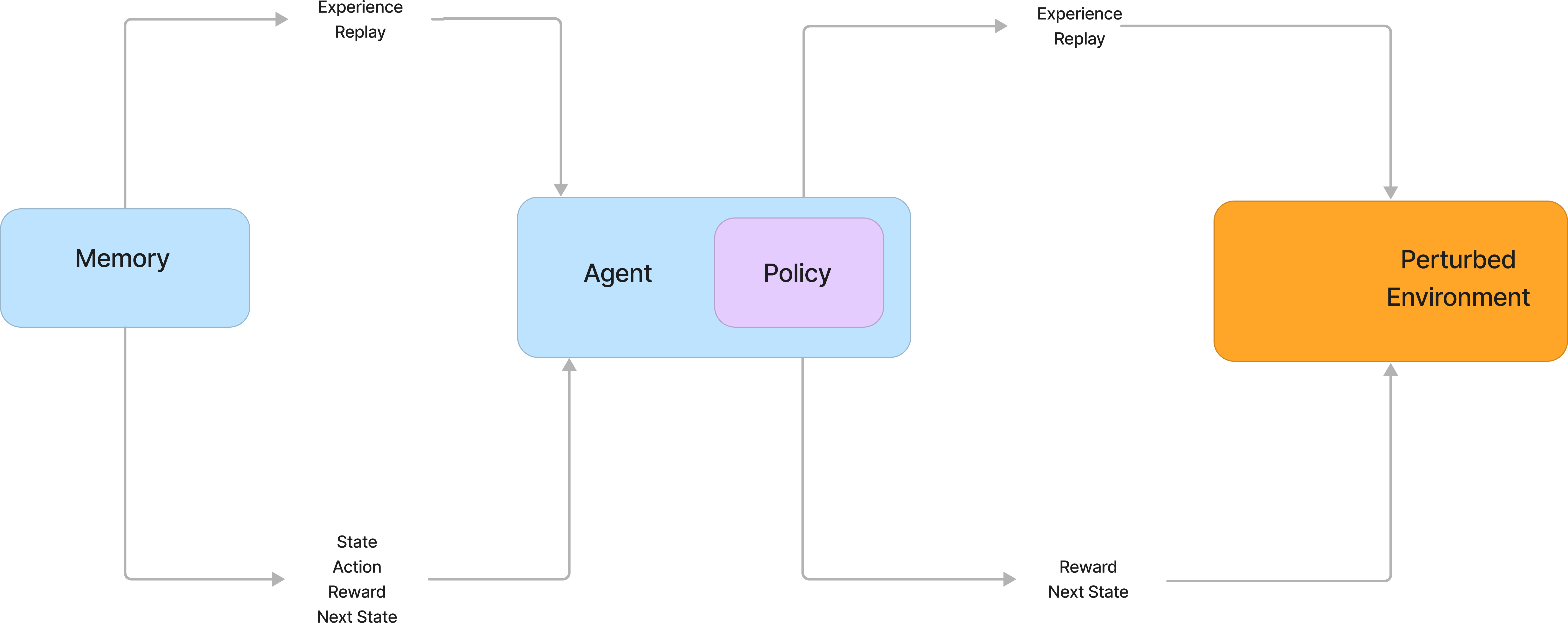} 
\caption{Adversarial Attacks in Reinforcement Learning \cite{Ilahi2021challenges}}\label{Fig1.1}
\end{figure}

Reinforcement learning is a process of self-optimization, where models evolve and improve, but the process is also threatened by AI security risks\cite{Ilahi2021challenges,Chen2022}. In the field of reinforcement learning, it is difficult to distinguish between the concepts of adversarial examples and AI backdoor, which are collectively referred to as "adversarial attacks". Based on attack paths, Kiourti et al.\cite{Kiourti2020trojdrl} categorize the attacks against reinforcement learning into environmental adversarial attacks, reward adversarial attacks, and adversarial policy attacks.

\begin{itemize}

\item \textbf{Environmental Adversarial Attacks}

Environment-based adversarial attacks are those that add perturbations to the environment perceived in reinforcement learning, thereby affecting the system's perception of state $s$, which in turn incorrectly matches the attacker's specified policy ${s}'$ to finally manipulate the decisions of the reinforcement learning system in a given state. In 2017, Huang et al.\cite{Huang2017adversarial} implemented an environmental adversarial attack on a reinforcement learning system by adding adversarial perturbations to the external environment image frames in reinforcement learning based on the white-box FGSM algorithm; in the same year, Lin et al.\cite{Lin2017tactics} proposed an optimized environmental adversarial attack method targeting at the best behavior in a specific state; Behzadan et al.\cite{Behzadan2017vulnerability} verified the transition of adversarial attacks among reinforcement learning models and thus proposed a transition-based black-box attack. In 2019, Xiao et al.\cite{Xiao2019characterizing} proposed a method for estimating model gradients based on frame consistency information, thus enabling the first adversarial black-box attack in reinforcement learning. In 2020, Kiourti et al.\cite{Kiourti2020trojdrl} proposed TrojDRL, which describes a reward-based adversarial attack as one that is performed by adding minute specific perturbations to the environmental state to enables $\hat{s} = s + \delta$ that eventually makes the behaviour given by the reinforcement learning model misunderstood, i.e.$ \hat{A}(s,m,\delta) \neq A(s,m,\delta) $. In the area of autonomous driving, Behzadan et al.\cite{Behzadan2019adversarial} in 2019 verified that in the environment of autonomous driving, using reward adversarial attacks, an attacker could cause a direct collision or malicious manipulation of the trajectory of the autonomous driving vehicles.

\item \textbf{Reward Adversarial Attacks}

If an attacker is able to maliciously tamper with some of the rewards, it may lead to the policy of the reinforcement learning system being manipulated by the attacker, thus posing a severe threat to the system. Kiourti et al.\cite{Kiourti2020trojdrl} validated a reward adversarial attack where an attacker would set the corresponding reward to 1 when its target state $s$ is reached, and otherwise set the reward to -1, to create a strong attack scenario. In 2022, Islam et al.\cite{Islam2022triggerless} proposed a reward adversarial attack applicable to the UAV environment.

\item \textbf{Adversarial Policy Attacks}

Unlike the reward adversarial attack approach, an adversarial policy attack does not need to tamper with the victim's reward or policy; instead, in an adversarial environment, the attacker quickly finds a policy to defeat the victim by analyzing the victim's policy or behavior, finding its vulnerabilities and exploiting them. Tretschk et al.\cite{Tretschk2018sequential} proposed the adversarial policy of Adversarial Transformer Networks (ATN). In 2020, Gleave et al.\cite{Gleave2019adversarial} proposed Adversarial Policies, in which an attacker generates targeted adversarial policies based on the behavior of the victim, producing seemingly random and uncoordinated behavior to defeat or disrupt the victim. Such policies are more successful in high-dimensional environments and have been validated in real eSports environments. In 2021, Wang et al.\cite{Wang2021backdoorl} synthesized reward adversarial attacks with adversarial policy attacks and proposed BackDooRK, which significantly improved the success rate of attackers in defeating their victims.

\end{itemize}


\textbf{Algorithm 11. Adversarial Policies}

For multiple (in the case of two) participants in a reinforcement learning environment, it is assumed that the victim's policy $\pi_{v}$ has been determined, and here the victim's policy determines its behaviour $a_{v} \sim \pi_{v}( \cdot \mid s ) $. And the attacker continuously optimizes its own policy $\pi_{\alpha}$ based on the victim's policy and behaviour. It then be described as in this Markov decision process $M_{a}=(S,A_{\alpha},T_{\alpha},{R}'_{\alpha})$, considered within state transition probabilities $T_{\alpha}$ and rewards $R_{\alpha}$, the integration of the victim policy $\pi_{v}$ yields

$$ T_{\alpha}(s,a_{\alpha}) = T(s,a_{\alpha,a_{v}}) $$ and

$$ {R}'_{\alpha}(s,a_{\alpha,{s}'}) = R_{\alpha}(s,a_{\alpha,a_{v},{s}'})  $$

The attacker finds an adversarial policy against the victim by optimizing the following loss function:
\begin{equation}
 \arg \max \sum_{t=0}^{\infty } \gamma^{t}R_{\alpha}(s_{t},a_{\alpha}^{t},s^{t+1} )   
\end{equation}

where it is subject to $s^{t+1} \sim T_{\alpha}(s^{t},a_{\alpha}^{t}) $ and $a_{\alpha} \sim \pi( \cdot \mid s^{t} )  $


\subsection{Summary}

This chapter introduces some technologies that may pose a security risk to the autonomous driving AI decision-making layer and briefly describes the technical principles. The main function of the sensor layer is to recognize the raw information collected by sensors, while the main function of the decision layer is to make driving decisions based on the perceived state of the environment. Each layer has its own autonomous driving function, and a threat to any of these layers could affect the overall safety of the autonomous driving system.

\section{Emerging Threats in Federated Learning-based Vehicular Internet of Things}

With the rapid development of smart vehicles, the vehicle is no longer an isolated single point, but increasingly a terminal worker in the pan-vehicle network. In many countries and regions around the world, the vehicular internet of things is already under rapid construction and its security based on traditional cyber security technologies has been widely studied\cite{Ren2019security}. However, with the development of emerging technologies such as arithmetic networks and privacy computing, new generation technologies such as deep learning technologies, edge computing and federated learning will be further integrated into the environment of the vehicular internet of things, giving rise to new business forms. Still, new technologies and new business forms also bring new security risks, and this chapter focuses on the new risks that they may bring.

It is well known that current artificial intelligence technology is a data-driven approach\cite{Lecun2015deep,Bengio2021deep,Zhang2020}. In AI applications, a large amount of information needs to be collected in advance as the training data. Traditional methods of data collection and information interaction face a number of limitations: First, traditional data collection and transmission are often inefficient. Secondly, traditional data collection often leads to invasion of user privacy. Therefore, Federated Learning (FL)\cite{Konevcny2016federated-1,Konevcny2016federated-2,Mcmahan2016federated} is a new distributed learning framework that does not require data collection by a central worker, providing a relatively more efficient and private way of interaction. In federated learning, each worker is trained with local data sets to obtain local gradients or weights through machine learning algorithms, and then uploads local gradients instead of local sensitive data, enabling knowledge interaction instead of data interaction. Federated learning has been widely used in the Internet, mobile terminals and other fields. At the same time, federated learning has also been introduced into the field of the vehicular internet of things \cite{Elbir2020federated,Posner2021federated,Zhang2021real,Du2020federated} and has become a future trend.

The federated learning environment offers more new attack methods\cite{Lyu2020threats,Mothukuri2021survey,Yin2021comprehensive}. With more workers participating in federated learning and the trustworthiness of each worker with the cloud difficult to guarantee, malicious workers in the vehicular IoT may attack the federated learning system in a variety of ways, while the privacy of the workers may also be at risk.

\subsection{Byzantine Attack on Federated Learning}

The Byzantine Attack in Federated Learning refers to workers attacked by a malicious Byzantine worker to construct harmful gradients which after aggregation will make the global model difficult to aggregate, thus making the system unusable or generating a global model with a malicious backdoor. Distinguished from traditional Data Posioning attacks, the above attacks are also known as Model Posioning attacks. In 2017, Blanchard et al.\cite{Blanchard2017machine} first proposed the Byzantine attack in a machine learning environment. The principle is that in round $t$, a non-Byzantine worker $p$ in federated learning will locally compute the unbiased estimate $V_{p}^{t}$ of its gradient $\bigtriangledown Q(x_{t})$ and  send it to the aggregation worker which according to some aggregation rule $F$, aggregates the received gradient estimates, then in round $t+1$, the weight of the global model is

$$ x_{t+1} = x_{t} - \gamma_{t} \cdot F(V_{1}^{t},...,V_{n}^{t}) $$

where $\gamma_{t}$ is the learning rate. While the malicious Byzantine worker cleverly constructs destructive local gradient estimates:

$$V_{n} = \frac{1}{\lambda_{n}} \cdot U - \sum^{n-1}_{i=1} \frac{\lambda_{i}}{\lambda_{n}} V_{i} $$

where $\gamma_{t}$ is the weight the gradient estimate of  worker $n$ at the time of aggregation. This will cause the global gradient to become any harmful gradient $U$ provided by the Byzantine worker after the final aggregation.

\begin{figure}[htb]
\centering
\includegraphics[width=7cm]{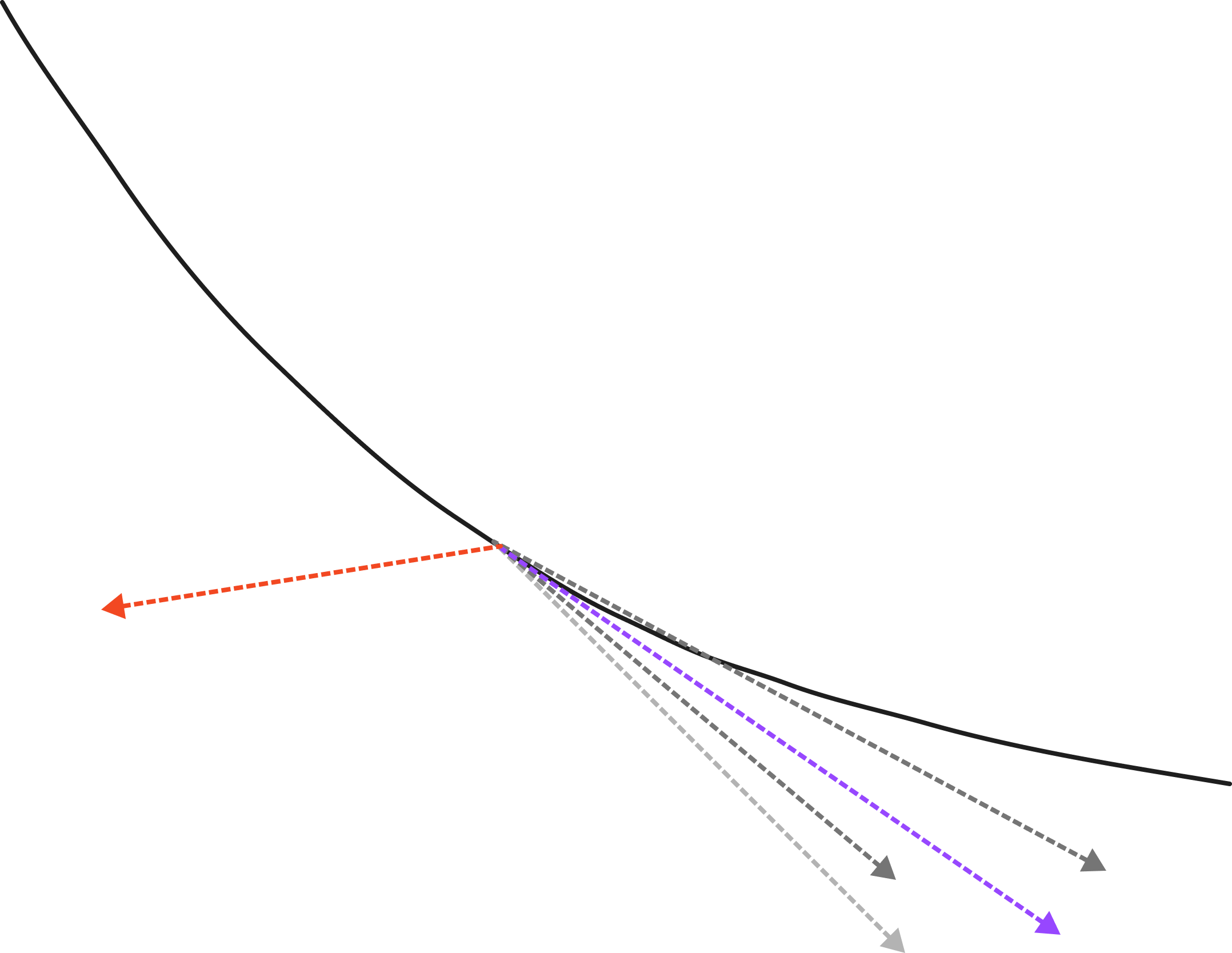} 
\caption{Schematic of Byzantine attack  \cite{Blanchard2017machine}. The black dashed line represents the gradient estimate for non-Byzantine workers, the blue solid line
represents the global gradient after normal aggregation, and the red dashed line represents the gradient estimate submitted by the Byzantine worker.}\label{Fig5.1}
\end{figure}

The above describes a Byzantine worker that  poisons in only some round of aggregation $t$, which is called a "single-shot attack". Generally, its limitations are as follows.

\begin{itemize}

\item \textbf{Attack's Capability Is Limited. } Attack just is a single worker or a single aggregation that has limited influence on its attacking power.

\item \textbf{Poor Concealment.} A single worker attack or a single-shot attack on an aggregator usually makes the state of the poisoned worker significantly different from a normal worker, and that leads to easily detected.

\item \textbf{Prone to Recession}. After multiple aggregations, the effect of poisoning a single worker during an aggregation round tends to fade, even does not continue to be effective.

\end{itemize}

Blanchard et al.\cite{Blanchard2017machine} also proposed the concept of Byzantine tolerance for measuring the robustness of federated learning models against Byzantine attacks. To improve Byzantine robustness, some scholars have proposed novel aggregation algorithms that can be used for federated learning\cite{Guerraoui2018hidden,Yin2018byzantine}.

To overcome these limitations, an attacker can adopt to \textbf{Repeated Attacks} or \textbf{Collusion Attacks}. Repeated Attacks mean that the attacker can perform poisoning in multiple rounds of aggregation; Collusion Attacks are joint poisoning of multiple Byzantine workers. Experiments show that repeated attacks and collusion attacks can enhance the capability of Byzantine attacks and can significantly improve the concealment and recession resistance of the attacks\cite{Bagdasaryan2020backdoor}. Xie et al.\cite{Xie2019dba} proposed a distributed backdoor attack that can be performed in a federated learning environment.


\textbf{Algorithm 12. Distributed Backdoor Attack} (DBA)\cite{Xie2019dba}

Distributed backdoor attacks provide a efficient way for multiple malicious workers in federated learning to conspire to an attack. The DBA algorithm takes advantage of the local data opacity in federated learning, with multiple malicious workers each adding more minute malicious perturbations in multiple rounds to improve concealment. The malicious permissions of each worker are optimally generated in the following manner.

\begin{equation}
\begin{split}
w^{*}_{i} = &\arg \max_{w_{i}} (\sum_{j \in s_{poi}^{i}}P[G^{t+1}(R(X_{i}^{j},\phi^{*}_{i})) = \tau; \gamma; I] \\ 
&+ \sum_{j \in S^{i}_{cln}} P [G^{t+1}(x_{j}^{i})=y_{j}^{i}]),  \forall i \in [M] 
\end{split}
\end{equation}

where $\phi^{*}_{i} =  \left \{ \phi, O(i)  \right \} $ represents the local poisoning policy of attacker $m_{i}$, and $\forall$ is the global trigger.


\begin{figure}[htb]
\centering
\includegraphics[width=\linewidth]{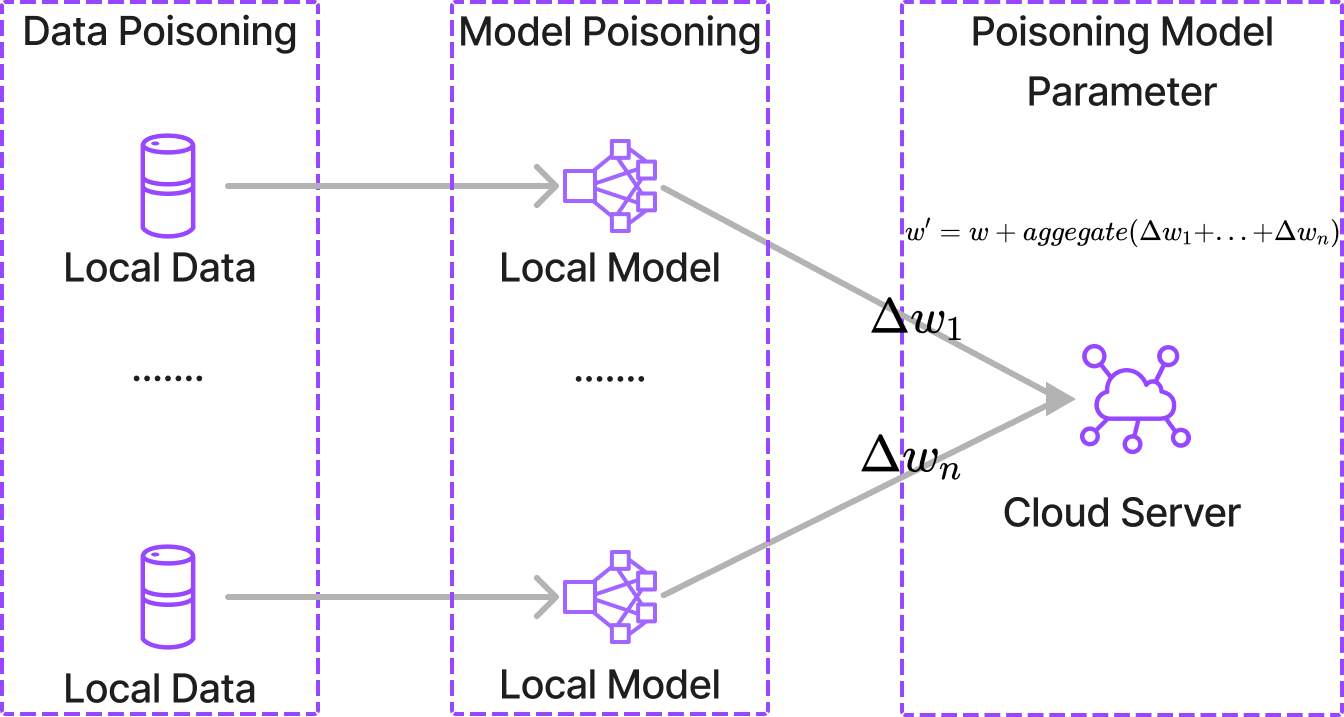} 
\caption{Schematic of Byzantine Attack \cite{Lyu2020threats}. The black dashed line denotes the gradient estimate for non-Byzantine workers, the blue solid line denotes the global gradient after normal aggregation, and the red dashed line denotes the gradient estimate submitted by the Byzantine worker. \cite{Lyu2020threats}}\label{Fig5.2}
\end{figure}

\subsection{Privacy Inference on Federated Learning}

The interior and exterior images of an autonomous driving vehicle may include sensitive information such as faces and license plate numbers. So local user data may reflect the user's location and trajectory, in-vehicle behavior and driving habits, which are also often considered sensitive, and therefore direct user data capture may lead to user privacy violations. One of the aims of Federated Learning is to avoid the direct leakage of sensitive user data, thereby achieving user privacy protection. The effects of user privacy in federated learning have received extensive attention and researches\cite{Kairouz2021advances,li2020federated,Geiping2020inverting,Al2019privacy, Lyu2020threats,Li2021survey,Zhou2021,Liu2020} . However, it shows that local gradients are highly correlated with the user's data and that data may still be inferred when local gradients are obtained. Techniques such as Model Inversion and Membership Privacy Inference may help infer local user data.

\begin{itemize}

\item Model Inversion
The model inversion method replaces the pixels in the initial random images one by one, then classifies the constructed images with the help of some model, and iteratively optimizes the constructed image based on the results of the classification, resulting in a constructed image that is highly similar to the target image. To speed up aggregation, model inversion algorithms mostly utilize the distribution of the target image as prior knowledge to participate in the optimization process. In 2015, Fredrikson et al.\cite{Fredrikson2015model} first proposed a model inversion attack; in 2016, Wu et al.\cite{Wu2016methodology} proposed a black-box model inversion attack algorithm; in 2017, Hitaj et al.\cite{Hitaj2017deep} proposed a model inversion algorithm based on adversarial generative network techniques, which achieved better reconstruction results, also known as "Adversarial Generative Network Reconstruction Attacks" (GAN Reconstruction Attacks). Mai et al.\cite{Mai2018reconstruction} extended model inversion to the field of face recognition to recover from face recognition features for face image data, proposing the Face Recovery Attack, and Razzhigaev et al.\cite{Razzhigaev2020black, Razzhigaev2021darker} continuously optimized the black box face recovery attack.

\item Membership Privacy Inference
Member privacy inference means that an attacker exploits the special feedback of the overfitting phenomenon and tries to infer whether the target data is in the training set or not. Typically, attackers first construct a Shadow Model similar to the target model; then uses the shadow model to generate training data; furthermore, used the data to train an Attack Model;  last, the attackers use the Attack Model to construct the complete attack process. In 2017, Shokri et al.\cite{Shokri2017membership} at Cornell University first proposed the member privacy inference attack. Once proposed, the member privacy inference method has been continuously researched and further optimized and improved\cite{Liu2020,Zhou2021,Hu2021membership}. In 2018, Yeom et al.\cite{Yeom2018privacy} analyzed in depth the relationship between overfitting and the risk of member privacy leakage. In 2019, Salem et al.\cite{Salem2019ml} incorporated a data transition attack method that reduces the attacker's reliance on background knowledge and makes the "shadow model" less necessary. Sablayrolles et al.\cite{Sablayrolles2019white} improved the membership privacy inference attack using a Bayesian optimization policy. Zhang et al.\cite{Zhang2021membership} extended the member privacy inference attack to the recommender system domain. Hui et al.\cite{Hui2021practical} proposed the BlINDMI algorithm, which first generates a certain amount of non-membership data, and then iteratively generates a comparison between non-membership data and membership data to improve the accuracy of membership privacy inference.

In a federated learning, the data distribution of workers is broadly similar. That providing more contextual knowledge and creating better conditions for malicious workers and the cloud to conduct member privacy inference attacks. In 2019, Nasr et al.\cite{Nasr2019comprehensive} validated the membership privacy inference attack risk in federated learning. In 2020, Chen et al.\cite{Chen2020beyond} further improved the success rate of member privacy inference attacks in federated learning using adversarial generative networks for data augmentation. In 2021, Hu et al.\cite{Hu2021source}  proposed Source Inference Attack for federated learning, which uses Bayesian methods to infer the training data of federated learning workers and in the same year, Gupta et al.\cite{Gupta2021membership} extended the traditional membership privacy inference attack from classification tasks to regression tasks, and also verified that their method is equally applicable to regression tasks in a federated learning environment.

\end{itemize}


\textbf{Algorithm 13. General Model Inversion}\cite{Fredrikson2015model}

Model inversion used be recover the training dataset that probability includes sensitive data. The general model inversion algorithm first computes each possible target feature vector $v$ for feature $x_{1}$ and then evaluates its probability of being correct. Also, since the class distribution of deep learning models tends to obey a Gaussian Distribution, so adapt a Gaussian function as a penalty function can accelerate the convergence. The algorithm can be described as follows.

\begin{algorithm}[h]
\caption{\textbf{13. Model Inversion}}\label{alg:cap}
\setcounter{AlgoLine}{0}
\KwInput{Target model $f$}
\KwOutput{An example of training data $x'$}
\LinesNumbered
\SetAlgoLined

initialized $x_{0}$ ;

 \ForEach  {feature $x_{i} \in$  the feature vector $X$} {
         
         \ForEach {the possible value $v \in x_{i}$ }{
         
                           $x' = {v,x_{2},x_{3},...,x_{n}}$
                           
                           $r_{v} \leftarrow  err(y,f(x')) \cdot \Pi_{i} p_{i}(x_{i}) $

          }
}
\end{algorithm}

\textbf{Algorithm 14. Membership Privacy Inference}\cite{Shokri2017membership}

Implementing a member privacy inference attack requires several processes, starting with the generation of data for training the shadow model. If the attacker has some background knowledge and possesses some homogeneous distribution data, it can be used directly for shadow model training. If the attacker does not have the appropriate background knowledge, data with a high confidence level can be selected as integrated training data by querying the target model. Each classification category $c$ will be initially recorded as $x$ randomly and iterated as follows: sequentially select the record data classified by the target model as $c$ with the maximum confidence level $y_{c}$ which is ensured to be greater than a certain threshold on $f$ , into the integrated data set. Once a record $x$ is selected into the integrated data set, randomly change the features as many as $k$ based on $x$ to generate a new record $X^{*}$. This is iterated until a certain amount of training data is generated. The second process is to generate a shadow model. Based on the generated training data set$D^{train}_{shadow_{i}}$, after training, shadow models $shadow_{i}$ will be generated. The third step is to train the attack model. Query the prediction vector of the records $(x,y) \in D_{shadow}^{train}$ in the training data set of shadow models $shadow_{i}$, then record $(y,\hat{y},in)$ can be generated. And calculate the prediction vector$\hat{y} = f_{shadow}^{i}(x)$ in the shadow model test data set $\forall(x,y) \in D_{shadow_{i}}^{test} $ then vector $\hat{y}=f_{shadow}^{i}(x)$ can be obtained. Next the two corresponding sets of vectors of each category $c$ are aggregated into the training data $D_{attack}^{train}$ of the attack model. On this basis, a classifier is trained to determine whether the data is included in the training data.


\subsection{Summary}

With the fusion and development of technologies, such as federated learning , edge computing and etc., the intelligent vehicular internet of things is gradually growing and becoming a future trend\cite{Nguyen2021deep}. However, the security and user privacy risks associated with federated learning and other technologies are also a concern. In the vehicular IoTs, the data distribution of each end is similar, providing attackers with certain background knowledge. Once a end is controlled by an attacker, the whole IoT networks may be subject to Byzantine attacks, and the risk of user privacy analysis is greatly increased. While enjoying the convenience offered by vehicular IoT and AI, we should not ignore the associated risks, but rather conduct relevant research and security protection.

\section{Conclusion}

Autonomous driving is a complex system based on artificial intelligence technology. A number of artificial intelligence applications, such as objective detection, segmentation, speech recognition and driving decision-making, play an important role in autonomous driving. Safety is a key concern in autonomous driving systems. AI security is crucial and directly affects autonomous driving system security, and leads to personal safety, which is far beyond traditional network security and basic software security. The novel technologies, such as AI,  bring emerging risks. 

This paper briefly introduces the AI technology route and AI functional modules in the autonomous driving system, and analyses the origins, development, and current appropriate AI security technologies for autonomous driving. Similar to general AI, autonomous driving is under threats of adversarial examples attacks, including AI backdoor attacks, model inversion and member privacy. 
Despite there are some defense methods that can be useful against these threats, ensuring safety for complex autonomous driving systems requires not only single-point defense techniques against certain threats but also building a complete trusted AI system\cite{Kaur2022trustworthy, Floridi2019establishing}, as following:

\begin{itemize}

\item  \textbf{Trustworthy AI evaluation system. }The safety of AI in autonomous driving system requires a complete evaluation and validation system, which covering data preparation, model training, model deployment, system application and other parts of the AI model life cycle. Especially, there are some noteworthy issues, including: AI adversarial robustness assessment, cross-domain data robustness assessment, model safety validation, training data safety validation, and data adversarial example detection.

\item  \textbf{Trustworthy AI Architecture.} Further more, we need to improve autonomous driving security from just reducing some specific threats to building Trustworthy AI Architecture. There is something beyond adversarial detecting need to do to build an AI architecture with human agency and oversight, robustness and safety, privacy and data governance, transparency, diversity, non-discrimination and fairness, societal and environmental well-being, and accountability.
\end{itemize}

To summarize, this paper appeals for attention and focus on emerging technologies such as artificial intelligence that bring new safety risks to autonomous driving systems. It is necessary that construct a safer and trusted autonomous driving system through the establishment of a trusted artificial intelligence technology system.

\section{Appendix A. Summary table} 

Here we present a table containing a summary of the adversarial examples algorithms as the foundation in this paper.

\begin{table*}[H]
\caption{Foundational Adversarial Examples Algorithms}
\label{table7.1}
\begin{center}
\begin{tabular}{ p{2.5cm}<{\centering} | p{4.5cm}<{\centering} | p{7.5cm}<{\centering}  }
 \hline
 \hline
 \rowcolor{lightgray}
 \textbf{Method} & \textbf{method} & \textbf{Major algorithms} \\ 
 \hline
 \hline
 
 \rowcolor{green!50!yellow!10}
 \multirow{3}{*}{White-Box} & 
 Gradient sign-based & 
 FGSM\cite{Goodfellow2014explaining}, BIM\cite{Kurakin2018adversarial}, MIM\cite{dong2018boosting},PGD\cite{madry2017towards} \\
 
 & 
 Optimization-based & 
 CW\cite{Carlini2017towards},  \\
 
 \rowcolor{green!50!yellow!10}
 & 
 Others & 
... \\

 \multirow{4}{*}{Black-Box} & 
 Transfer-Based & 
 DIM\cite{xie2019improving}, TI\cite{dong2019evading} \\
 
 \rowcolor{green!50!yellow!10}
 &
 Approximate Gradient &
 BPDA\cite{athalye2018obfuscated},EoT\cite{Athalye2018synthesizing} \\
 
 & 
 Score-based & 
 ZOO\cite{chen2017zoo}, NES\cite{ilyas2018black}, SPSA\cite{uesato2018adversarial}, $N$Attack\cite{li2019nattack} \\

 \rowcolor{green!50!yellow!10} 
 & 
 Decision-based & 
 Boundary Attack\cite{brendel2017decision}, Evolutionary Attack\cite{dong2019efficient} \\ 

 \hline
 \hline
 
\end{tabular} 
\end{center}
\end{table*}









\bibliography{8Reference.bib}







\end{document}